\documentclass[article,onecolumn]{aastex62}
\makeatletter
\let\@dates\relax
\makeatother

\usepackage{subfigure}
\usepackage{graphicx}
\usepackage{multirow}
\usepackage{url}
\usepackage{hyperref}
\usepackage{lineno}
\newcommand{\RoboBA}{{\tt RoboBA}}
\newcommand{\dol}{{\tt DoL}}
\newcommand{\balrog}{{\tt BALROG}}
\newcommand{\Swift}{{\it Swift}}

\shorttitle{GBM Localizations}
\shortauthors{Goldstein et al.}

\begin{document}

%%Title
\title{Evaluation of Automated Fermi GBM Localizations of Gamma-ray Bursts}

%%Authors
\author[0000-0002-0587-7042]{A.~Goldstein}
\affiliation{Science and Technology Institute, Universities Space Research Association, Huntsville, AL 35805, USA}

\author{C.~Fletcher}
\affiliation{Science and Technology Institute, Universities Space Research Association, Huntsville, AL 35805, USA}

\author{P.~Veres}
\affiliation{Center for Space Plasma and Aeronomic Research, University of Alabama in Huntsville, 320 Sparkman Drive, Huntsville, AL 35899, USA}

\author{M.~S.~Briggs}
\affiliation{Space Science Department, University of Alabama in Huntsville, 320 Sparkman Drive, Huntsville, AL 35899, USA}

\author{W.~H.~Cleveland}
\affiliation{Science and Technology Institute, Universities Space Research Association, Huntsville, AL 35805, USA}

\author{M.~H.~Gibby}
\affiliation{Jacobs Technology, Inc., Huntsville, AL 35805, USA}

\author{C.~M.~Hui}
\affiliation{Astrophysics Office, ST12, NASA/Marshall Space Flight Center, Huntsville, AL 35812, USA}

\author{E.~Bissaldi}
\affiliation{Istituto Nazionale di Fisica Nucleare, Sezione di Bari, I-70126 Bari, Italy}

\author{E.~Burns}\thanks{NASA Postdoctoral Fellow}
\affiliation{Goddard Space Flight Center, Greenbelt, MD 20771, USA}

\author{R.~Hamburg}
\affiliation{Space Science Department, University of Alabama in Huntsville, 320 Sparkman Drive, Huntsville, AL 35899, USA}

\author{A.~von~Kienlin}
\affiliation{Max-Planck-Institut f\"{u}r extraterrestrische Physik, Giessenbachstrasse 1, 85748 Garching, Germany}

\author{D.~Kocevski}
\affiliation{Astrophysics Office, ST12, NASA/Marshall Space Flight Center, Huntsville, AL 35812, USA}

\author{B.~Mailyan}
\affiliation{Center for Space Plasma and Aeronomic Research, University of Alabama in Huntsville, 320 Sparkman Drive, Huntsville, AL 35899, USA}

\author[0000-0002-0380-0041]{C.~Malacaria}\thanks{NASA Postdoctoral Fellow}
\affiliation{Astrophysics Office, ST12, NASA/Marshall Space Flight Center, Huntsville, AL 35812, USA}

\author{W.~S.~Paciesas}
\affiliation{Science and Technology Institute, Universities Space Research Association, Huntsville, AL 35805, USA}

\author{O.~J.~Roberts}
\affiliation{Science and Technology Institute, Universities Space Research Association, Huntsville,
AL 35805, USA}

\author{C.~A.~Wilson-Hodge}
\affiliation{Astrophysics Office, ST12, NASA/Marshall Space Flight Center, Huntsville, AL 35812, USA}

\begin{abstract}
The capability of the Fermi Gamma-ray Burst Monitor (GBM) to localize gamma-ray bursts (GRBs) is evaluated for two different automated algorithms: the GBM Team's \RoboBA\ algorithm and the independently developed \balrog\ algorithm. Through a systematic study utilizing over 500 GRBs with known locations from instruments like Swift and the Fermi LAT, we directly compare the effectiveness of, and accurately estimate the systematic uncertainty for, both algorithms. We show simple adjustments to the GBM Team's \RoboBA, in operation since early 2016, yields significant improvement in the systematic uncertainty, removing the long tail identified in the systematic, and improves the overall accuracy. The systematic uncertainty for the updated \RoboBA\ localizations is $1.8^\circ$ for 52\% of GRBs and $4.1^\circ$ for the remaining 48\%. Both from public reporting by \balrog\ and our systematic study, we find the systematic uncertainty of $1-2^\circ$ quoted in circulars for bright GRBs is an underestimate of the true magnitude of the systematic, which we find to be $2.7^\circ$ for 74\% of GRBs and $33^\circ$ for the remaining 26\%.  We show that, once the systematic uncertainty is considered, the \RoboBA\ 90\% localization confidence regions can be more than an order of magnitude smaller in area than those produced by \balrog.
\end{abstract}

%% Introduction
\section{Introduction}
To date, the Fermi Gamma-ray Burst Monitor~\citep[GBM;][]{Meegan09} is one of the most prolific instruments for the prompt detection of gamma-ray bursts (GRBs).  Its onboard trigger algorithms detect $\sim240$ GRBs per year, and for each of those, the Fermi GBM team sends public alerts within seconds of detection and transfers quicklook trigger data with a latency of $\sim 10$ minutes.  This prompt public delivery of data has enabled community scientists rapid access to GBM data for triggered GRBs and has lead to several follow-up campaigns of GBM-detected GRBs as well as prompt correlation analysis between GBM triggers and triggers by other GRB-detecting instruments.  Additionally, the full science data from GBM, both for onboard triggers and continuous data, is downlinked to the ground every few hours and is available publicly promptly after processing of the raw data into calibrated files.  A top priority of the GBM Team is to provide such a rapid delivery of all data (typically $\ll 1$ day), enabling and encouraging community scientists to use GBM data for all manner of investigations, especially in the realm of time-critical observations and analysis.  

In addition to the delivery of data for public use, one of the primary tasks of the Fermi GBM Team is to provide prompt localizations of GRBs detected by the GBM flight software.  Unlike the Fermi Large Area Telescope (LAT) or the Neil Gehrels \Swift\ Observatory, GBM does not have the capability of imaging or of reconstructing arrival directions of individual photons, and therefore it cannot consistently localize a signal on the sky to sub-degree precision.  Instead, the localization method uses the relative rates of scintillation detectors oriented in different directions
to estimate the source location, as pioneered by the KONUS experiments \citep{Mazets}. The GBM localization algorithm is based on that used for BATSE on the Compton Gamma-ray Observatory~\citep{Pendleton99} and provides localizations for GBM with a statistical precision down to 1 degree and a typical localization region of several degrees in radius~\citep{Connaughton15}.  What GBM lacks in precise localization power, however, it makes up with its all-sky monitoring capability, fine-time resolution, broad energy coverage, and fine spectral discrimination.  

Localization of GRBs and other transients observed by GBM are of prime importance in the era of multi-messenger and time-domain astronomy.  The benchmark example is the joint detection of GW170817 and GRB 170817A by LIGO/Virgo and Fermi GBM~\citep{170817Joint, 170817Team}.  These two detections of a binary neutron star merger in gravitational waves and gamma rays were entirely independent detections of the same event, and the consistent localization from both detections were an important piece in determining the high confidence of the connection between the two.  Other examples include serendipitous observations of the sky or follow-up by other instruments.  GRB 161228B was found by the intermediate Palomar Transient Factory (iPTF) to be a GRB counterpart to a Type IC broad-line Supernova, as discovered due to the localization agreement between the GRB and the supernova location~\citep{Corsi17}.  iPTF and now its upgrade, the Zwicky Transient Facility, has followed-up GBM GRBs and discovered optical counterparts by utilizing only the GBM localizations~\citep{Singer13, Singer15, Coughlin19}.  Other wide-field telescopes have begun follow-up as well, including MASTER~\citep{Lipunov16}, and GOTO~\citep[e.g.][]{Mong19,Ulaczyk19}.  Finally, searches for counterparts to detections in other instruments, from radio to gamma-ray and including gravitational-wave and neutrino instruments, utilize GBM GRB localizations~\citep{Abbott17, Cunningham19, Ho19, Abbott19}.

To ensure that these and future efforts are productive, an accurate estimate of the GBM localization systematic uncertainty is a requirement to prevent reporting over-confident localizations that may result in false associations or lead to ruling out real associations.  To that end, we determine with high fidelity the systematic uncertainty for two GBM localization algorithms: the GBM Team's official automated system, termed the \RoboBA, and the localization algorithm it uses, the \dol~\citep{Connaughton15}; and the independently developed \balrog\ algorithm~\citep{Burgess17}. Both of these algorithms use the real-time trigger data (TRIGDAT) provided by the onboard GBM flight software.  The GBM flight software monitors the detector rates, and when it detects a statistically significant rate increase, it ``triggers,'' entering a mode that includes the production of special data.   The trigger data includes onboard calculated localizations, which are constrained by using only a pre-burst background average and the severely limited onboard computational resources, and thus are less accurate but useful in the onboard classification of triggers. The TRIGDAT data also contains lightcurve data for quick look analysis.  The TRIGDAT time history data  extends from $\sim$130 s before the trigger to $\sim$470 s after the trigger, and is provided on timescales ranging from 64~ms to 8.192~s, with the shorter timescales concentrated around the trigger time. To transmit the TRIGDAT data in real time, the Fermi spacecraft initiates an unscheduled transmission with a NASA Tracking and Data Relay (TDRS) using the TDRS Demand Access Service.   This service allows on-demand access, but provides only very low bandwidth. Collecting 470 s of data post-trigger, and transmitting the TRIGDAT takes 10 minutes -- this dominates the latency of the \RoboBA.  The post-trigger data allows ground software to create background models using pre- and post-GRB data, which can result in significantly improved background estimates and thus improved localizations.

We first describe the \RoboBA\ and \balrog\ algorithms in Sections~\ref{sec:RoboBA} and~\ref{sec:balrog}, respectively, and then we discuss the methodology and samples of known GRB locations used to estimate the systematic uncertainty in Section~\ref{sec:MethodAndSamples}.  Finally, we discuss the overall results and implications in Section~\ref{sec:Discussion}.

% The RoboBA description
\section{The Fermi GBM \RoboBA}\label{sec:RoboBA}
The \RoboBA, operational since early 2016, is a set of algorithms developed to run autonomously to replace the Human-in-the-Loop (HitL) processing for most GBM GRB triggers.  HitL processing requires burst advocates (BAs) to be on-call at all times and ready to promptly handle the processing of GBM triggers.  Due to the increasing interest and importance in GBM-detected GRBs, localizations of GRBs are desirable as soon as possible.  HitL processing had a median 1--2 hour delay in sending out final localizations, so the fully-automated \RoboBA\ was developed to provide localizations for GRBs with accuracy on the order of human BA processing and a latency within 10 minutes after trigger (within seconds after complete receipt of TRIGDAT) and also reports in the GCN notice whether the GRB is likely to be a short or long duration GRB.  The \RoboBA\ has an automatic processing success rate of $\gtrsim85\%$, with most failures due to dropped data packets in the real-time communication stream from the spacecraft.  Since the first implementation in early 2016, there have been a few updates to the \RoboBA, namely an update in early 2018 to provide localizations in full-sky HEALPix~\citep{Healpix} FITS files and an update in May 2019 to automatically send a GCN circular in addition to the machine-readable GCN notice~\citep{AutoCirc19}. We present here another update, implemented in September 2019, that further optimizes the performance of the \RoboBA, resulting in overall improved localization accuracy and a smaller systematic uncertainty~\citep{NewRobo19}.  We first provide an overview of the \RoboBA\ algorithm and then detail the most recent improvement in Section~\ref{sec:RoboBA_Improvements}.

\subsection{Background Estimation}
The \RoboBA, implemented in the GBM Burst Alert Pipeline (BAP), receives the real-time data as it arrives, sequences the data in the correct time order, and constructs the lightcurve.  If the real-time trigger data is sufficiently complete, the \RoboBA\ performs a series of background fits, utilizing a two-pass recursive non-parameteric regression.  The non-parametric regression can accommodate a variety of non-stationary background conditions that are exhibited in GBM data, and it can handle the very limited amount of trigger data (usually $\sim170$ time bins total).  Since GRB lighcurves exhibit a wide variety of durations and morphologies, the regression is performed recursively, first fitting the full set of data, then removing bins from the background that exceed a predefined S/N threshold.  The regression is repeated on the remaining time bins and those above the S/N threshold are removed until the recursive process converges, typically within 4 iterations.  The first pass of this fit is performed on the sum of all energy channels for each detector to boost signal statistics.  The second pass then performs the fitting process on each individual energy channel with the bins exceeding the S/N threshold already removed.  Potential failures in the background fitting process are identified and reported to the human BA for inspection and manual localization.  Failures in the background fitting typically only occur during extreme background variations and TRIGDAT truncation, which can occur when GBM triggers on a GRB as Fermi is entering or exiting the South Atlantic Anomaly.

\subsection{Signal Identification}
Once the background has been estimated, the signal needs to be identified.  Because the primary goal is to perform a localization, and because the current GBM localization algorithm uses the signal in 50--300 keV energy range, we perform the following procedure utilizing only the data in that energy range.  First, only considering the longer timescales in the TRIGDAT (8.192 s and 1.024 s), a calculation of each bin containing signal is performed assuming a mixture model of background+signal, with the signal model represented as an exponential function of signal counts during the trigger data window. This effectively assumes the background model is correct.  Ideally, both background model and signal model would be simultaneously estimated, however the small number of time bins in TRIGDAT does not permit reliable convergence of such estimation for most GRBs.  The prior assumption used in calculating the probability that a bin contains signal is approximately the fraction of bins above a predefined S/N threshold.  The probability is calculated for each time bin in each detector and then the probabilities are combined across detectors such that incoherent background fluctuations in different detectors produce lower probabilities and coherent signal across detectors produce higher probabilities.  Bins with a final probability $> 0.89$ are considered as bins containing significant signal useful for localization. If the duration of the bins with significant signal is less than 2.1 s in duration, or if there is no identifiable signal on the longer timescales, then the shorter duration (64 ms and 256 ms) bins are considered and analyzed for the presence of a signal.  If there is still no identifiable signal, then the \RoboBA\ cannot proceed and the human BA is notified for a manual localization.  This type of failure typically only occurs for extremely weak triggers or trigger with durations $< 64$ ms (the GBM flight software can trigger on shorter timescales than are delivered in the TRIGDAT).

\subsection{Signal Selection}
Once the bins with likely signal presence have been identified, a selection can be determined for localization purposes.  The GBM localization method is best performed when a contiguous segment of data has been identified and that segment is no longer than 30 s in duration.  Signal selections longer than 30 s will typically result in a larger localization uncertainty due to the fact that the spacecraft is moving significantly relative to the GRB location, and the localization calculation is performed in the spacecraft frame. The \RoboBA\ identifies the segment of the signal that contains the largest number of counts in order to minimize the statistical uncertainty of the localization, but will truncate the segment if it extends beyond 30 s in duration. The truncation is performed at either end of the segment, choosing to truncate the end that contains the fewest counts.

\clearpage

\subsection{Localization and BAP Processing}
Once the \RoboBA\ has estimated the background and identified the signal for localization, the inputs are sent to the GBM localization algorithm called the \dol\ (Daughter of LOCBURST), which is an updated version of the localization code used by the BATSE Team to localize GRBs~\citep{Pendleton99}.  The \dol\ assumes three spectral templates, selected to span the range of GRB spectra, and folds these template spectra through the NaI detector responses over a grid of points on the sky in the spacecraft frame.  This produces model counts in each detector that is compared to the observed counts above background in each detector, which is done specifically in the 50--300 keV range for GRBs.  Ultimately, the directional nature of the NaI responses and the blockage by the spacecraft body enable the localization of GRBs, since a specific region of the sky will produce a ratio of counts in the different detectors that match the observed ratio of counts. \citet{Connaughton15} provides the full description of the algorithm and the initial estimate of the GBM localization systematic uncertainty related to this algorithm.

\begin{figure}
	\begin{center}
		\includegraphics[scale=0.5]{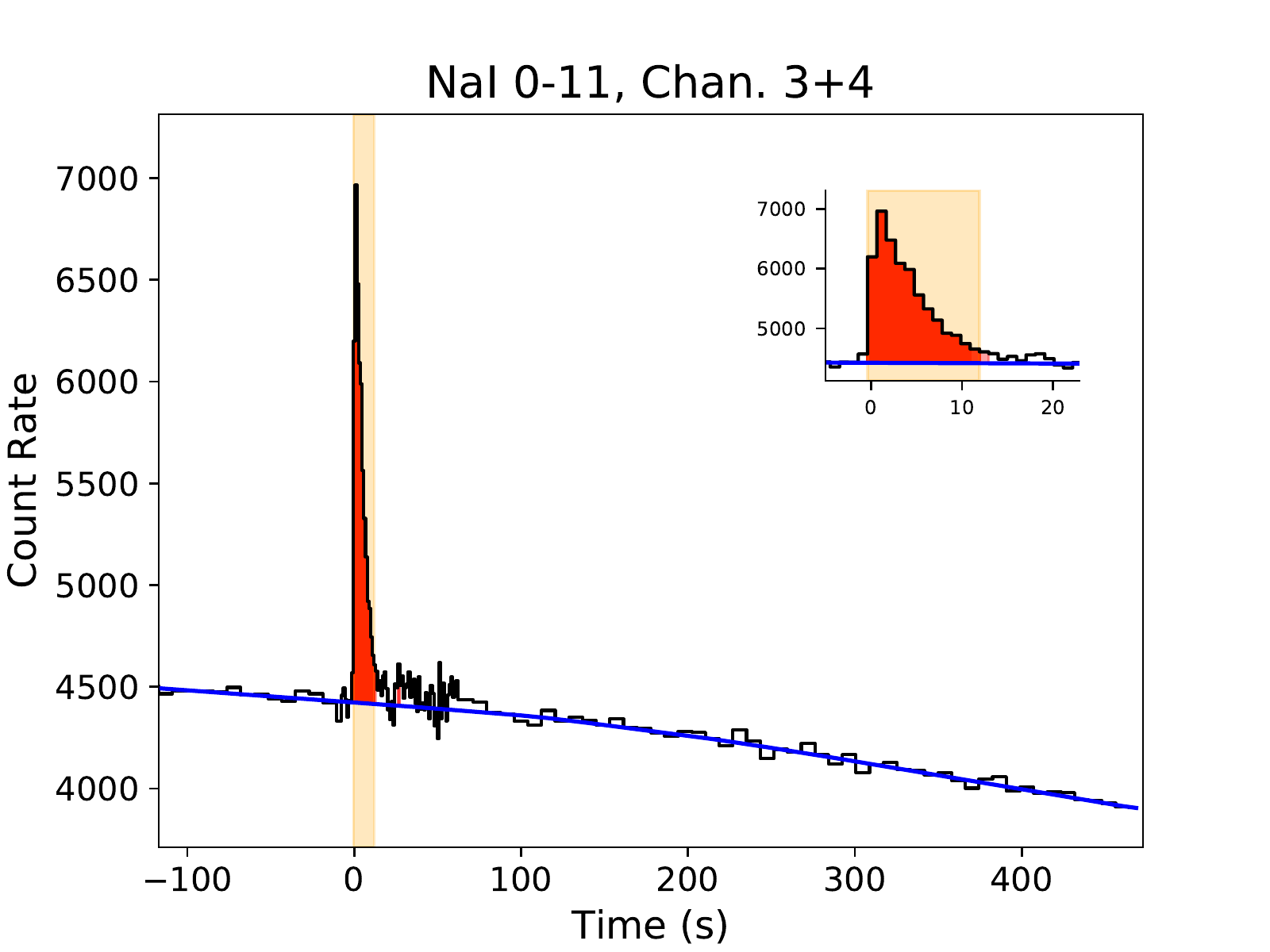}
		\includegraphics[scale=0.5]{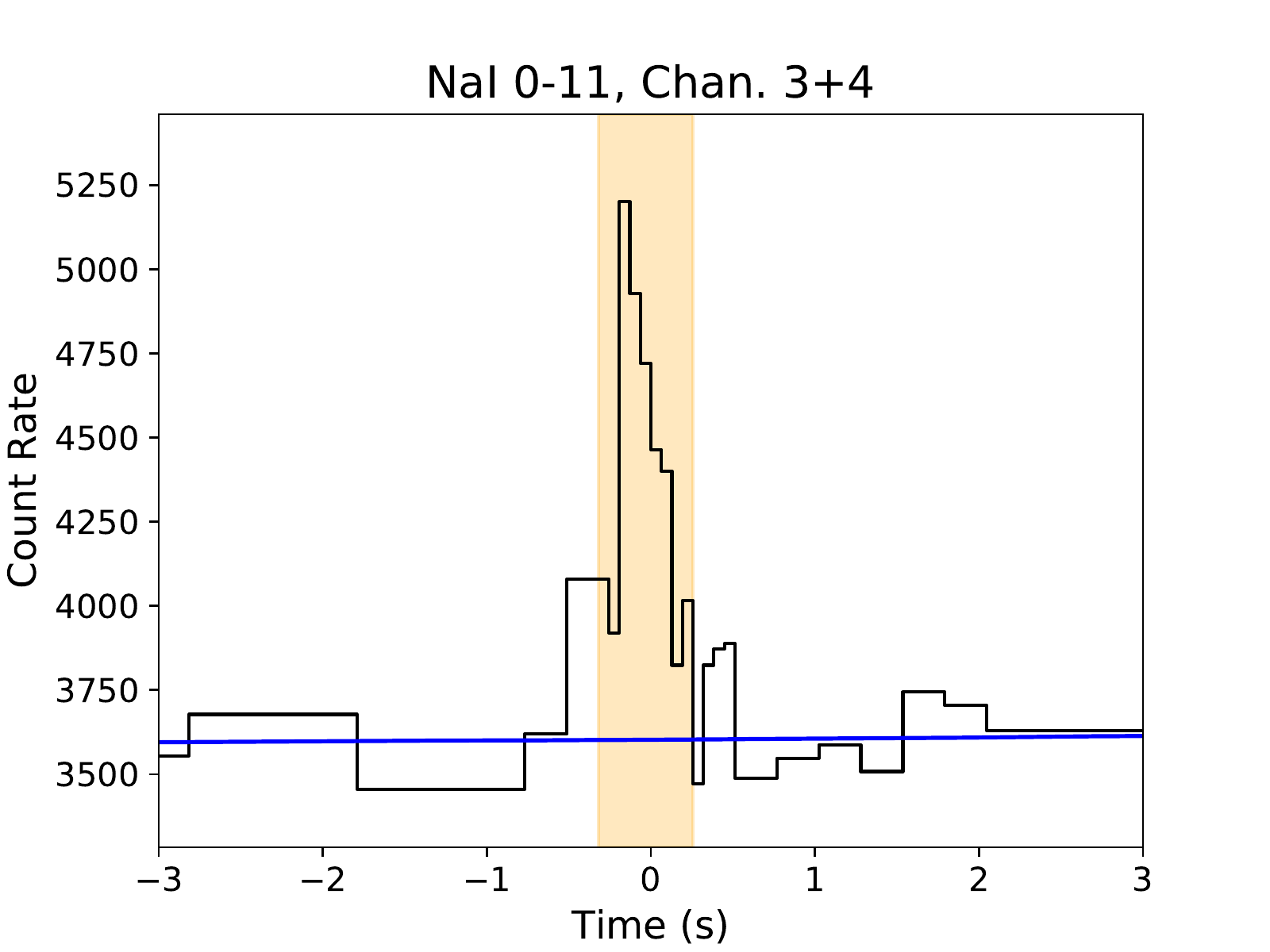}
	\end{center}
\caption{\RoboBA\ processing of  {\it [Left]} Fermi GBM long GRB 190222537 and {\it [Right]} short GRB 170817A.  Shown is the sum of the data from all NaI detectors in the 50--300 keV energy range.  The blue line is the background estimate, and the red filled region for the long GRB is identified as containing significant signal.  The orange selection shows the segment of data chosen to perform the localization analysis.
\label{RoboBA_LC}}
\end{figure}

Once the \dol\ has provided the localization based on the inputs from the \RoboBA, the BAP submits the GCN Notice\footnote{\url{https://gcn.gsfc.nasa.gov/fermi_grbs.html}} and Circular for the \RoboBA\ localization, and localization products, including the full-sky HEALPix map of the localization incorporating the estimated systematic uncertainty model, are uploaded to the HEASARC\footnote{\url{https://heasarc.gsfc.nasa.gov/FTP/fermi/data/gbm/bursts/}}.  The entire process of the \RoboBA\ from receipt of the initial trigger packet to the sending of the GCN notice and circular takes 10 minutes, with virtually all of that time devoted to waiting for complete receipt of the data.  Once the TRIGDAT packets have been received, the \RoboBA\ completes its operations on the order of 1 second, and the \dol\ operates in $< 10 $ s on a single CPU.  In addition to the public notices and maps, diagnostic lightcurves are produced for the human BA to verify the processing by the \RoboBA, and examples are shown in Figure~\ref{RoboBA_LC}.

\subsection{Algorithm Improvements}\label{sec:RoboBA_Improvements}
The \RoboBA\ algorithm was formulated and began operation prior to the production of the full-sky HEALPix maps, which provide a complete evaluation of the localization accuracy, precision, and systematics.  Here we detail small improvements made to the \RoboBA\ algorithm in order to make it more accurate and robust, and the evaluation of the systematic for these improvements are detailed in Section~\ref{sec:RoboResults}. One concern to be addressed by improving the \RoboBA\ is the increased systematic uncertainty compared to the HitL, which the GBM Team accepted as a trade-off for the gain of prompt reporting (10 minutes instead of 1--2 hours).  However, the most important property of the localization posterior is the total area once the systematic is incorporated, therefore only comparing the systematic uncertainty between the HitL and \RoboBA\ localizations does not provide an accurate representation of the capability of the \RoboBA.  Therefore, we have undergone some review of the \RoboBA\ algorithm, testing small changes to the parameters controlling its operation and testing changes to the three spectral templates used in the \dol, to provide an improvement in both accuracy and reliability.  

The \dol, since beginning of GBM operations, has used three spectral templates for localization, all of the form of a Band function~\citep{Band93}.  These are termed as `hard', `normal,' and `soft.'  Initially, testing focused on replacing the hard template, which has an unphysically hard high-energy power law, since it does not represent any known GRB spectrum.  Testing on variations of the Band function yielded some improvement, however the most obvious improvement in localization accuracy resulted from replacing all Band function templates with cut-off power laws (commonly referred to as a Comptonized function, parameterized with $E_{\rm peak}$).  The Comptonized parameters were then selected so that a majority of the GRB spectra would be represented by the normal spectrum, and the hard and soft spectra would represent the tails of the GRB spectral distribution. The parameters for the old Band function templates and the new Comptonized function templates are shown in Table~\ref{tbl:templates}.

\begin{table}[!h]
\begin{center}
\begin{tabular}{l | c c c}
\hline
Templates & Low-energy Index & High-energy Index & $E_{\rm peak}$ (keV) \\
\hline
Old Hard & 0.0 & -1.5 & 1000\\
Old Normal & -1.0 & -2.3 & 230\\
Old Soft & -2.0 & -3.4 & 70\\
\hline
\hline
New Hard & -0.25 & - & 1000\\
New Normal & -1.15 & - & 350\\
New Soft & -1.95 & - & 50\\
\hline
\end{tabular}
\caption{The parameters of the old and new spectral templates used in the \dol\ for localization.  The old templates were Band functions, and the new templates are Comptonized functions.
\label{tbl:templates}}
\end{center}
\end{table}

The \RoboBA\ adjustments included improvements to the background fitting stability and robustness, allowing for a larger number of missing packets in the realtime TRIGDAT. The threshold to allow bins to be considered for localization was lowered from 89\% to 75\%, and the maximum duration of localization signal was shortened from 30~s to 15~s. Finally, the signal threshold for the short timescales was lowered from $3\sigma$ to $1.5\sigma$, reducing the failure rate for localizing weak short GRBs.  All of these changes were relatively minor---an adjustment of a parameter in most cases---and included no additional complexity.  The changes were tested on a sub-sample of GRBs with known locations (see Section~\ref{sec:sample}) and validated on the remaining sample to ascertain an improvement in both localization accuracy and robustness. Indeed, after implementing these changes in the \RoboBA\ and \dol, the localization accuracy and posterior sky area are now improved over even that of the HitL processing, with no added complexity or run-time.  Additionally, the estimated failure rate is $<5\%$, reduced from $\sim15\%$, largely due to ability to handle an increased number of dropped data packets.  A few failures result from the misclassification of the GRB as `long' or `short' which determines the timescales of the TRIGDAT that are used for localization.  By a simple comparison of the classification provided by the \RoboBA\ to the final $\rm T_{90}=2 $~s split, the \RoboBA\ correctly classifies long GRBs 95\% of the time and correctly classifies short GRBs 92\% of the time.  Not all misclassifications result in a localization failure, but they may require human follow-up to provide a more accurate localization. The remaining failures are dominated by severe TRIGDAT truncation and extreme background variability around entry and exit of the SAA.  We detail the result of the improvements on the localizations in Section~\ref{sec:RoboResults}.

% The BALROG description
\section{The \balrog\ algorithm}\label{sec:balrog}
The \balrog\ algorithm~\citep{Burgess17} is designed to provide a significant improvement for the GBM localization of GRBs by jointly fitting the GRB spectrum and position on the sky. In principle, fitting the spectrum allows a closer match to the true spectrum of the GRB and therefore should provide more accurate localizations with smaller systematic uncertainties.  In practice, the \balrog\ algorithm performs a Monte Carlo sampling both on the sky and in spectral parameter space, using a parameterized spectral function. Per guidance from~\citet{Berlato19}, three different spectral functions are tried (a power law, an exponentially cut-off power law, and a Band function), and the localization using the best-fit spectrum is selected.  Unlike the \RoboBA, which can run and produce a localization in $< 10$ s on a single modern CPU, the \balrog\ algorithm requires the GBM detector responses to be generated for each Monte Carlo sample and therefore takes anywhere from an hour to a few hours to run using the three spectral functions on similar hardware.  This implies the need for significant computing resources, such as a computing cluster, to produce localizations with a latency of several minutes.

For validation, the \balrog\ algorithm was tested specifically on bright GRBs with known locations, while weaker GRBs were not studied due to the assumption that systematic uncertainty is only a dominant contribution to the total uncertainty when the GRB is bright.  Based on a subset of 69 localizations of bright GRBs originally presented in~\citet{Connaughton15}, \balrog\ was shown to produce localizations with smaller angular offsets and significantly reduce the localization uncertainty, thereby making GBM localizations of GRBs both more accurate and more precise for most GRBs~\citep{Berlato19}. A slightly larger sample of GRBs (105) was used to estimate any remaining systematic uncertainty for \balrog. \citet{Berlato19} determined that the remaining systematic was between $1-2^\circ$, with GRBs arriving along the direction of the Fermi solar panels tending to have the larger systematic than those arriving on the sides of the spacecraft where the GBM detectors were oriented.  This remaining systematic was not rigorously or statistically estimated, but instead it was inferred visually and was intended only for an approximate comparison with the official GBM team localizations, not for use in assessing the true systematic.  However, beginning on March 12, 2019, \balrog\ localizations relying on the GBM Team's production of TRIGDAT began to be produced shortly after the trigger data became public~\citep{Greiner19}, utilizing the $1-2^\circ$ systematic.

\subsection{Public Reporting}
Since the beginning of public reporting of \balrog\ localizations through August 2019, there have been 23 GRBs with precise or well-constrained localizations from other instruments for which \balrog\ produced a public localization.  While GCN circulars were not sent for all \balrog\ localizations, the localizations are available publicly on the MPE website and are accessible through the links provided in the GCN circulars~\citep{Greiner19}.  The fact that the \balrog\ localizations are now produced with a small latency after the GBM Team localization allows for a near-realtime, blind, and unbiased comparison of how the localization algorithms fare.  For the GBM Team localizations, GCN notices are sent for every localization and the localization files are hosted through the Fermi Science Support Center, and therefore the localizations have a public record. For the \balrog\ localizations, a record is available for those that have corresponding GCN circulars.  In Table~\ref{tbl:realtime}, we show the comparison between the \RoboBA\ and \balrog\ localizations for these 23 GRBs.  When GCN circulars are available for \balrog\ localizations, we use the information contained within, including the link to the HEALPix sky maps, while for the remaining GRBs we use the information publicly reported on their website.  To establish a public record of the localization at, or near, the time that they were produced, we have archived\footnote{\url{http://web.archive.org}} the relevant \balrog\ webpages and to prevent confusion in case there are future updates to the localization. We note that care must be taken when retrieving the \balrog\ HEALPix sky map from the link in the GCN Notice, since it may retrieve the localization using TTE data instead of the TRIGDAT.  In addition to the properties reported in Table~\ref{tbl:realtime}, we include in the Appendix a sky map comparison of the \RoboBA\ and \balrog\ localizations for this sample.

\begin{table}[!h]
\begin{tabular}{c | c c | c  c | l}
\hline
GRB & \multicolumn{2}{c|}{\balrog} & \multicolumn{2}{c|}{GBM Team} & Localizing Instrument \\
 & Offset (deg) & Confidence Level  & Offset (deg) & Confidence Level & (GCN) \\
\hline
190320A & 2.47\footnote{\url{https://web.archive.org/web/20190620011653/https://grb.mpe.mpg.de/grb/GRB190320052/?data_version=0}} &  0.04 & 7.90  & 0.79 & Swift XRT (23977)\\
190324A & 6.92$^*$ & $1-(2\times10^{-16})$ & 1.10 & 0.08 & Swift UVOT (24014)\\
190331A & 2.84\footnote{\url{https://web.archive.org/web/20190620012114/https://grb.mpe.mpg.de/grb/GRB190331093/?data_version=0}} & 0.41 & 10.8 & 0.61 & Swift BAT (24030)\\
190415A & 4.25$^*$ & 0.994 & 2.81 & 0.44 & IPN box$^\dag$ (24128)\\
190422A & 18.95\footnote{\url{https://web.archive.org/web/20190620222328/https://grb.mpe.mpg.de/grb/GRB190422957/?data_version=0}} & 1.00 & 3.34 & 0.43 & Swift XRT (24151)\\
190427A & 48.73\footnote{\url{https://web.archive.org/web/20190427134111/https://grb.mpe.mpg.de/grb/GRB190427190/?data_version=0}} & 0.58 & 4.62 & 0.45 & Swift BAT (24261)\\
190511A & 7.12\footnote{\url{https://web.archive.org/web/20190511121640/https://grb.mpe.mpg.de/grb/GRB190511302/}} & 0.84 & 4.71 & 0.76 & Swift UVOT (24494)\\
190512A & 36.63\footnote{\url{https://web.archive.org/web/20190512171703/https://grb.mpe.mpg.de/grb/GRB190512611/?data_version=0}} & 0.98 & 3.71 & 0.24 & Swift BAT (24520)\\
190515A & 130.9\footnote{\url{https://web.archive.org/web/20190620005451/https://grb.mpe.mpg.de/grb/GRB190515190/?data_version=0}} & $1-(3.5\times10^{-5})$ & 10.0 & 0.68 & Fermi LAT (24560)\\
190519A & 5.78\footnote{\url{https://web.archive.org/web/20190620005805/https://grb.mpe.mpg.de/grb/GRB190519309/?data_version=0}} & 0.94 & 3.29 & 0.56 & Swift XRT (24597)\\
190530A & 0.64$^*$ & 0.45 & 3.34 & 0.46 & Swift UVOT (24703)\\
190531B & 2.94$^*$ & 0.66 & 4.05 & 0.52 & Swift XRT (24706)\\
190606A & 66.85\footnote{\url{https://web.archive.org/web/20190620013103/https://grb.mpe.mpg.de/grb/GRB190606080/?data_version=0}} & 0.995 & 6.74 & 0.93 & IPN box$^\dag$ (24765)\\
190613A & 6.98$^*$ & 0.04 & 3.44 & 0.43 & Swift UVOT (24815)\\
190613B & 4.41$^*$ & 0.73 & 6.99 & 0.71 & Swift UVOT (24817)\\
190719C & 2.93\footnote{\url{https://web.archive.org/web/20190719154128/https://grb.mpe.mpg.de/grb/GRB190719624/?data_version=0}} & 0.23 & 9.17 & 0.88 & Swift XRT (25125)\\
190727B & 1.97$^*$ & 0.83 & 4.65 & 0.76 & Swift BAT (25211)\\
190731A & 4.12\footnote{\url{https://web.archive.org/web/20190801113429/https://grb.mpe.mpg.de/grb/GRB190731943/}} & $1-(4\times10^{-4})$ & 2.79 & 0.37 & Swift XRT (25244)\\
190805B & 4.33$^*$ & $0.75$ & 2.44 & 0.26 & IPN box$^\dag$ (25316)\\
190821A & 12.1\footnote{\url{https://web.archive.org/web/20190821185532/https://grb.mpe.mpg.de/grb/GRB190821716/?data_version=0}} & 0.98 & 5.60 & 0.31 & Swift XRT (25436)\\
190824A & 2.65\footnote{\url{https://web.archive.org/web/20190828132304/https://grb.mpe.mpg.de/grb/GRB190824616/?data_version=0}} & 0.40 & 5.76 & 0.76 & Swift XRT (25466)\\
190828B & 12.6\footnote{\url{https://web.archive.org/web/20190828131909/https://grb.mpe.mpg.de/grb/GRB190828542/?data_version=0}} & 0.21 & 5.41 & 0.80 & Swift XRT (25521)\\
190829A & 2.32\footnote{\url{https://web.archive.org/web/20190829202611/https://grb.mpe.mpg.de/grb/GRB190829830/?data_version=0}} & 0.92 & 2.13 & 0.40 & Swift XRT (25567)\\
\hline
\end{tabular}
\caption{The list of GRBs with known or well-constrained locations for which the \balrog\ localization has been performed in near-real-time.  The angular offset is calculated from the peak posterior density, and the confidence level is where the true location of the GRB is relative to the give localization posterior, including the prescribed systematic.\\
$^{*}$ Reported in GCN: 190324A (23994), 190415A (24123), 190530A (24677), 190531B (24696), 190613A (24800), 190613B (24808), 190727B (25178), 190805B (25272) \\
$^{\dag}$ Calculated from center of box\\
\label{tbl:realtime}}
\end{table}

\begin{figure}[h]
	\begin{center}
		\includegraphics[scale=0.55]{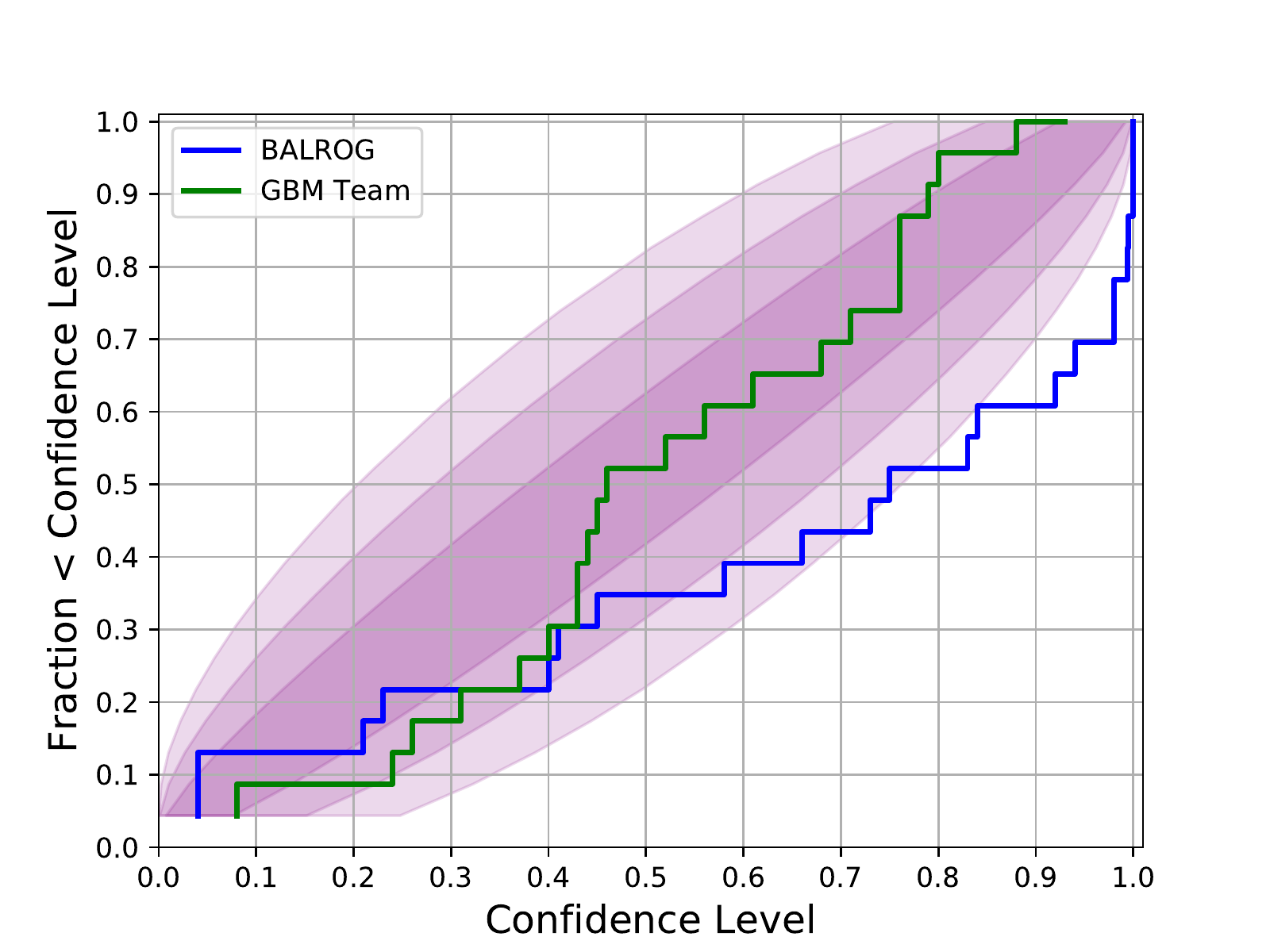}
		\includegraphics[scale=0.55]{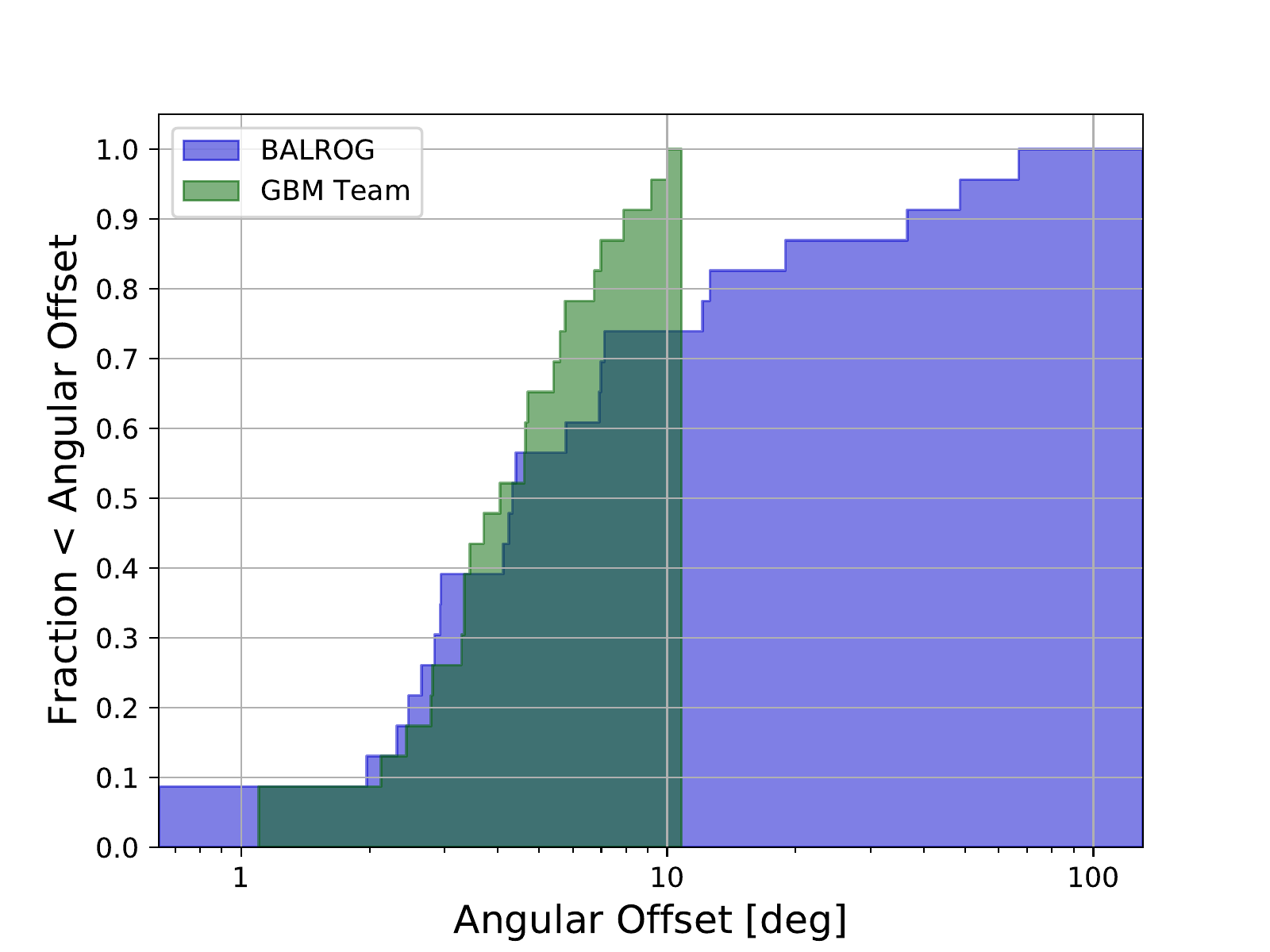}
	\end{center}
\caption{{\it [Left]} Cumulative distribution of the confidence level at which each of the known locations exist for the \balrog\ and \RoboBA\ localizations.  The purple bands represent the 1, 2, and 3$\sigma$ confidence regions for the expected equivalence of a perfectly calibrated localization uncertainty given a sample size of 23 GRBs. {\it [Right]} The cumulative distribution of angular offsets between the true location and the maximum density of the posteriors for \RoboBA\ and \balrog.\label{fig:realtime}}
\end{figure}

Table~\ref{tbl:realtime} lists the angular offsets of the peak probability density of each localization compared to the known location and the confidence level at which the true location lies, based on the posterior contained in the HEALPix files.  Note that for both \RoboBA\ and \balrog, the systematic uncertainty has already been incorporated, and in the case of \balrog, we use the systematic quoted in the GCN circular or the corresponding webpage.  Of note is that the \RoboBA\ localizations have a smaller offset from the true location for 15 out of the 23 GRBs, which is displayed in Figure~\ref{fig:realtime}, and that the \balrog\ excludes the true location at $\gtrsim 3\sigma$ for several GRBs, even though we include the prescribed systematic uncertainty. Figure~\ref{fig:realtime} also shows the probability--probability plot comparing the calibrations of the \RoboBA\ and \balrog\ localizations to expectations.  For perfectly calibrated statistical and systematic uncertainties, the cumulative distribution of true locations within the GBM localization posterior confidence levels should approach that of a uniform distribution.  That is, there should be a one-to-one relationship between the number of true locations that lie within a given confidence interval and the value of that confidence interval. For a finite sample size, especially a small sample size of 23, statistical fluctuations can cause significant deviation of the calibrated distribution away from the perfect assumption.  Therefore, we sample from a uniform distribution of size 23 and empirically determine the 1, 2, and 3$\sigma$ (Gaussian equivalent) confidence intervals.  We show the cumulative distributions for \RoboBA\ and \balrog\ in comparison to these confidence intervals.  While the localizations produced by the \RoboBA\ keeps within $2\sigma$ of a well-calibrated uncertainty, the \balrog\ localizations deviate significantly beyond $3\sigma$, and in fact are inconsistent with a well-calibrated uncertainty at  $>5\sigma$, indicating that the $1-2^\circ$ systematic claimed for \balrog\ is an underestimate of the true remaining systematic uncertainty.  This underestimate is potentially problematic because any follow-up undertaken utilizing \balrog\ localizations has a good chance at searching an incorrect part of the sky.
\clearpage

\subsection{Archival Testing}
Due to the small sample size of known locations since \balrog\ began public reporting, we installed and ran the \balrog\ algorithm on a larger sample of archival GRBs with known locations. Following the \balrog\ tutorial\footnote{\url{https://github.com/mpe-heg}} and the steps and best practices outlined in~\citet{Berlato19}, and with the desire to replicate the operational parameters of the real-time \balrog, we created a script to automatically process the necessary \balrog\ functions to localize a large number of GRBs with known locations.  The entire process was performed blindly with respect to the true location for each GRB, as is done for the \RoboBA. However, we note that initialization of the \balrog\ algorithm requires two additional pieces of information beyond what are available to \RoboBA: a determination of the signal timescale and a selection for the set of detectors which are most likely to provide the best localization.

Unlike the \RoboBA, which provides a classification on the type of GRB and therefore determines the timescale of data to use, the \balrog\ does not provide any automated identification of signal or signal timescale.  This implies that either several selections must be attempted and evaluated based on some criteria, or that visual inspection is required to evaluate the appropriate duration of signal to choose.  Due to computational considerations, we chose to identify the difference between long and short GRBs visually and select a signal interval of 8 s for long GRBs and a signal interval of 1--2 s for short GRBs. Once the signal selection had been performed, the background was then fit with a polynomial using a range of 100 seconds starting 20 seconds before the peak time bin and 50 seconds after, excluding the signal, to minimize source contamination in the background fit. 

The guidance provided in~\citet{Berlato19} indicates that not all GBM detectors should be used by \balrog\ for localization in most cases, but instead only detectors within a certain angle from the true location should be used, or detectors from the side that is consistent with the arrival direction of the GRB should be used.  
Since, in a blind test, we cannot know {\it a priori} the true arrival direction of the GRB, we choose to use the 6 NaI detectors and BGO detector that are on the same side of the spacecraft as the detector with the highest count rate.  We note that there are cases when the brightest detector does not necessarily indicate the general arrival direction of the GRB, such as in a case of high atmospheric scattering, but we choose this as our method so as to keep as close as possible to the operation of \balrog\ as outlined in~\citet{Berlato19}.

Once the signal had been identified, the background fit, and the appropriate detectors selected, we initialized \balrog\ with a set of reasonable initialization parameters, and it performed the localization of the signal using three spectral forms: a Band function, Comptonized function, and a power law.  As outlined in~\citet{Burgess17}, the spectrum with the lowest $log(z)$ (the marginal log-likelihood) was chosen as the preferred spectrum, and the resulting RA and Dec chains from \balrog\ were formatted into a HEALPix FITS map. Several \balrog\ localizations failed to converge, at a rate consistent with what is produced by the automated \balrog\ pipeline (see Appendix 1 for some examples).  In additional testing of the algorithm, we compared our results to those in~\citet{Berlato19}, and while for many GRBs we could reproduce their results, there were several GRBs for which we could not reproduce their results unless manual selections of the signal were made until the localizations were consistent.

\section{Localization Systematics Estimation}\label{sec:MethodAndSamples}

\subsection{Methodology}
The initial systematic uncertainty estimation for GBM localizations was described in detail in~\citet{Connaughton15}. The method for evaluating the systematic uncertainty assumed that the localization statistical uncertainty was a von Mises-Fisher distribution (Gaussian distribution on a sphere). A Gaussian-equivalent 1$\sigma$ radius was calculated by determining the area of the $1\sigma$ region defined by the $\chi^2$ statistic produced from evaluating the localization fitness over a $1^\circ$ resolution grid on the sky.  Various models for the systematic uncertainty were compared with a Bayesian technique, optimizing the parameters of each model using the log-likelihood and comparing models by odds ratios.   The probability of each localization was determined, and included in the likelihood, by using the distance of the GBM localization from a point source localization from another instrument or distance from an IPN~\citep{Hurley13} localization annulus.  This technique was originally developed to estimate the systematic uncertainty of BATSE GRB localizations \citep{Briggs99}. This method found that there was evidence for a two-component systematic, termed a `core+tail' model, which is a sum of two von Mises-Fisher distributions. There was also evidence that the systematic was different for two distinct slices in spacecraft azimuth, indicating directions in the spacecraft frame where there is a smaller systematic uncertainty.  In general~\citet{Connaughton15} found that the `core' contribution to the model ($\sim$88--92\% contribution) has a systematic defined by a 1$\sigma$ radius of 2.3--4.2 degrees and the `tail' contribution has a radius of 13.2--15.3 degrees. Note that since the core+tail model is an empirical estimation of the systematic that is not motivated by known contributors, one cannot predict {\it a priori} which GRB localization will be represented by the `core' component of the systematic or the `tail.'  Therefore, the systematic uncertainty is a mixture of the two components and is convolved with the statistical uncertainty to produce the final calibrated uncertainty. 

The assumed circular approximation for the localization uncertainty, however, is not accurate in many cases.  In practice, the statistical 
uncertainty of the localization can be quite complex, affected by the blockage of the detector fields by parts of the spacecraft, scattering of photons from the spacecraft, and backscattering of photons from the atmosphere.  Imperfect knowledge of the spacecraft mass model, the atmospheric scattering model, and the true spectrum of the source (as well as its inherent spectral evolution) all contribute to the systematic uncertainty, and each of these contributions are difficult to isolate, deconvolve, and correct.  We can evaluate the overall effectiveness of the systematic model, and thereby the Gaussian assumption, by performing a convolution of the systematic model with the statistical uncertainty and producing a probability-probability (P--P) plot, as was shown in Figure~\ref{fig:realtime}.

The P--P plot is a widely-used tool to investigate the similarity of two cumulative distributions.  In the context of measuring the 
contribution of systematic uncertainty to the estimation of a set of random variables, the P--P plot is useful in comparing the 
ensemble of proposed posteriors to the expected fraction of true values within specified confidence levels in the frequentist 
interpretation.   For example, if we calculate the localization posterior for a sample of GRBs of which we know the 
true location, the P--P plot allows us to determine if 50\% of the true locations fall inside the 50\% confidence region, 90\% of the 
true locations fall inside the 90\% confidence region, etc.  A unity line is often shown to establish the expected comparison, and a 
deficit in comparison to that line indicates that the posterior, on average, underestimates the uncertainty, while an excess over 
that line indicates an overestimate of the uncertainty.  In short, the P--P plot indicates the integrity of an uncertainty calibration.

Since the P--P plot is a measure of the calibration of the posterior, continuous along all confidence values, it can be used to 
estimate a systematic component of the posterior when a statistical estimate of the posterior underestimates the true uncertainty.  
This is useful whether or not the sources of systematic uncertainty are known, and whether or not several sources can be easily 
disentangled and modeled.  Generally, if one has a model for the systematic which has some free parameters, to calibrate the 
systematic against the P--P plot, the binomial likelihood function must be maximized:
\begin{equation}
	L(p|n, y)  \propto p^y (1-p)^{(n-y)},
\end{equation}
where $p$ is the binomial probability, $n$ is the number of trials in a given experiment, and $y$ is the number of ``successes'' in 
that experiment.  The binomial log-likelihood of $n$ events with a given ``success rate'' and expected probability along the P-P 
plot is:
\begin{equation}
	\ln{L} \propto \sum_{i=0}^n y_i \ln{p_i} + (n-y_i) \ln(1-p_i),
\end{equation}
and in this context we define success rate as the number of true locations within a given confidence level. In terms of the P--P plot, at any point along the cumulative distribution, one is measuring the number of ``successes'' given the data and comparing it to the expected number of successes from the model of the uncertainty.  To estimate the parameters of a systematic uncertainty model, we  maximize the binomial log-likelihood, which cannot be solved analytically in this situation.  Since the statistical uncertainty for a GBM localization can have a wide variety of morphologies without a standard analytical description, we must choose an optimization algorithm that will determine the parameter values that maximize the likelihood.

We choose to use the Nelder-Mead simplex algorithm~\citep{NelderMead} to minimize the negative log-likelihood by convolving a candidate systematic model with the statistical posterior for each GRB in the sample.  The fitting process proceeds as:
\begin{enumerate}
\item Initialize the systematic model with a guess parameter vector;
\item Convolve each GRB localization posterior with the systematic model;
\item Construct the cumulative distribution of the fraction of true locations within the confidence regions;
\item Calculate the negative binomial log-likelihood and evaluate the Nelder-Mead stopping criteria;
\item If the Nelder-Mead stopping criteria hasn't been satisfied, the Nelder-Mead algorithm formulates a new parameter vector and 
steps 2--4 are repeated until convergence.
\end{enumerate}

Once the fitting process has concluded, the minimum of the negative log-likelihood has been found. By assuming that likelihood surface in the region local to the minimum is approximately Gaussian, the covariance matrix of the fit can be estimated by calculating the inverse of the Fisher Information Matrix, $I(\theta)$, for the log-likelihood, which is the negative Hessian of the log-likelihood as a function of the systematic model parameters, $\theta$:
\begin{equation}
	\Sigma = [I(\theta)_{i,j}]^{-1} = E\biggl[-\frac{\partial^2}{\partial\theta_i \partial\theta_j} \ln{L(X;\theta)}\biggr| \theta \biggr]^{-1}.
	\label{eq:covar}
\end{equation}
In the case of the GBM localization posteriors, since the likelihood depends on the statistical uncertainty, which has no analytical 
form, we use finite differences to estimate the elements of the Hessian.

\subsection{Sample}\label{sec:sample}
To evaluate the efficacy of the current and updated \RoboBA\ localization systematic and the \balrog\ systematic, we use a sample of 516 GRBs with locations known to a 1-$\sigma$ precision of $\leq 1^\circ$ as determined by other instruments.  This sample represents a selection of GRBs spanning from the beginning of GBM operations in July 2008 through March 2019, and was gathered from the online GBM GRB catalog\footnote{\url{https://heasarc.gsfc.nasa.gov/W3Browse/fermi/fermigbrst.html}}.   This represents a factor $> 2.5$ larger sample compared to the constrained locations used in~\citet{Connaughton15} and a factor $\sim5$ larger sample compared to~\citet{Berlato19}, and combined with our improvements in methodology, we can create an increasingly robust estimation of the localization systematic uncertainty for both algorithms.

For both the \RoboBA\ and the \balrog\, we ran this sample of GRBs through automated processing blindly, without considering the known location.  In the case of the \RoboBA, the algorithm as implemented in the BAP was run on a desktop computer over all GRBs in the sample, meanwhile, we utilized the Alabama Supercomputer\footnote{\url{https://www.asc.edu/}} to run the \balrog\ algorithm so that the localizations could be performed in a tractable amount of time.  In both cases there were a few failures of the algorithms to identify and localize signals. The \RoboBA\ identified 11 failures out of the sample of 516, therefore we exclude those failures from consideration.  For \balrog, 17 failures were immediately identified that prevented the algorithm from completing, therefore we exclude those GRBs from the \balrog\ sample.  Additionally, several \balrog\ localizations would not converge, and no identifiable cause was found other than sensitivity to initial conditions.  We exclude these 103 GRBs from consideration for \balrog, leaving a sample of 396 GRBs.  We use the full sample of successful localizations to estimate the systematic for each  algorithm, and we use the sub-sample of joint successes for the head-to-head comparisons.

\subsection{\RoboBA\ Results}\label{sec:RoboResults}
The \RoboBA\ localizations were fit using the prescribed methods, first starting with a single Gaussian representing the systematic for all GRBs, and then with the `core+tail' model. In all cases, the single Gaussian systematic is not sufficient to accurately describe the systematic, therefore the preferred model for the systematic is the `core+tail' model.  For the original \RoboBA\ algorithm, the systematic was calibrated with a smaller sample of GRBs under the Gaussian assumption used in~\citet{Connaughton15}.  In initial testing of the \RoboBA, it was determined that the systematic was noticeably larger for short GRBs than for long GRBs, and since the majority of GRBs detected by GBM are long, the systematic underestimated the true uncertainty for localizations of short GRBs.  Therefore, we estimated the systematic for long and short GRBs (as classified by \RoboBA) separately, and show those models in Table~\ref{tbl:RoboBA_fits} in comparison to the HitL systematic.  While the \RoboBA\ localizations do not exhibit a $>10^\circ$ tail for long GRBs as was estimated for the HitL localization, the short GRBs localizations are contaminated with a small fraction of very bad localizations, producing a tail that extends out to $\sim 30^\circ$. Improvements to the \RoboBA\ were specifically targeted at solving this tail, along with generally reducing the overall systematic.  

\begin{table}[!h]
\begin{center}
\begin{tabular}{ c | c c c c }
\hline
\tablecolumns{5}
Model & Sample & Core ($^\circ$) & Core \% & Tail ($^\circ$) \\
\hline
\hline
\multirow{3}{*}{HitL} & All GRBs & $3.7\pm0.2$ & $90\pm4$ & $14\pm3$\\
\cline{2-5} 
                      & Az: $292.5-67.5^\circ$ \& $112.5-247.5^\circ$ & $4.2\pm0.3$ & $92\pm4$ & $15\pm4$\\
                      & Az: $67.5-112.5^\circ$ \& $247.5-292.5^\circ$ & $2.3\pm0.4$ & $88\pm6$ & $13\pm4$\\

\hline
\multirow{2}{*}{Original \RoboBA} & Long GRBs & $2.6\pm0.1$ & $65\pm4$ & $6.0\pm1.0$\\
                                  & Short GRBs & $3.6\pm0.1$ & $98\pm1$ & $29.6\pm15.6$\\
\hline
\hline
\multirow{6}{*}{Updated \RoboBA} & All GRBs & $1.81\pm 0.02$ & $51.7\pm 1.2$ & $4.07\pm 0.05$\\
\cline{2-5} 
                                 & Long GRBs & $1.86\pm 0.02$ & $57.9\pm 1.2$ & $4.14\pm 0.06$\\
                                 & Short GRBs & $2.55\pm 0.08$ & $39.0\pm 1.2$ & $4.43\pm 0.16$\\
\cline{2-5} 
                                 & Hard Spectrum & $2.38\pm 0.07$ & $52.8\pm 0.9$ & $4.97\pm 0.12$\\
                                 & Normal Spectrum & $1.94\pm 0.04$ & $62.4\pm 0.6$ & $3.44\pm 0.09$\\
                                 & Soft Spectrum & $1.40\pm 0.03$ & $40.4\pm 0.6$ & $4.05\pm 0.06$\\
\hline
\end{tabular}
\caption{The systematic uncertainty models for the HitL, original, and updated \RoboBA.
\label{tbl:RoboBA_fits}}
\end{center}
\end{table}

 Considering the updated \RoboBA, we are afforded the opportunity to test different hypotheses on the sources of the systematic with such a large sample of GRBs.  As was shown in~\citet{Connaughton15} and later in~\citet{Berlato19}, there is evidence for an azimuthal dependence of the systematic in spacecraft coordinates, which would imply an angular-dependent systematic in the GBM detector responses.  Since identifying and correcting a putative systematic in the response is outside the scope of this work, and since the arrival direction of a GRB is not known {\it a priori} in useful circumstances, modeling the systematic as a function of arrival direction provides a poor predictive solution.  Instead, we aim to further model the systematic based on the characteristics of the \RoboBA\ so that those characteristics can be used to leverage a more accurate and optimal estimate of the systematic on a per-GRB basis.  Specifically, we investigate the systematic resulting from the different preferred spectral templates, since a source of systematic can be the imperfect assumption of the spectrum for the GRB~\citep{Burgess17, Berlato19}.  We also investigate the systematic for what the \RoboBA\ algorithm determines are long and short GRBs, as was done for the original \RoboBA.  Since the \RoboBA\ makes a classification of the GRB, and the signal identification and selection methodology is somewhat different between what the \RoboBA\ determines is a long or short GRB, there can be different contributions of systematic uncertainty due to the methodology.

\begin{figure}
	\begin{center}
		\includegraphics[scale=0.55]{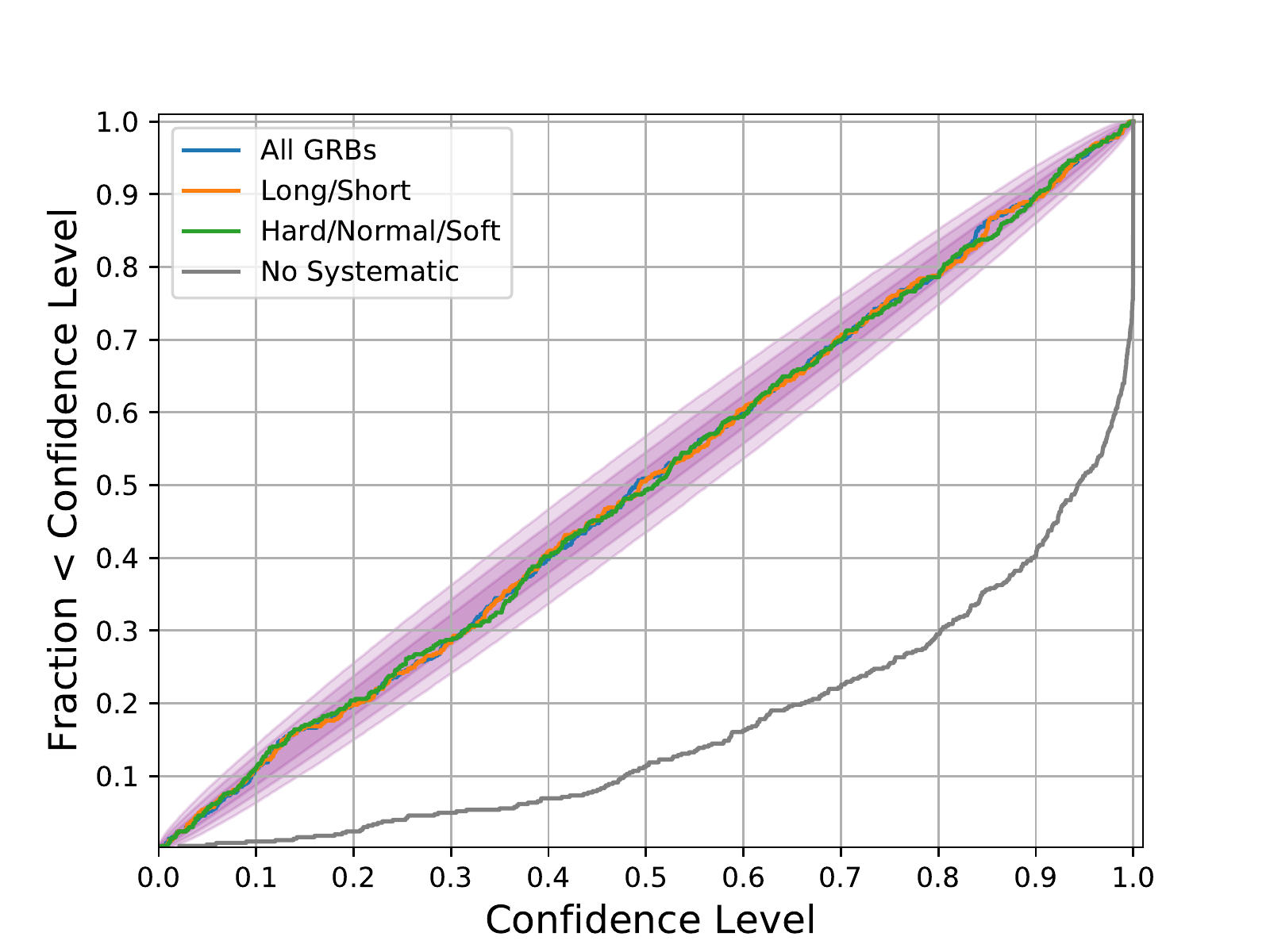}
		\includegraphics[scale=0.55]{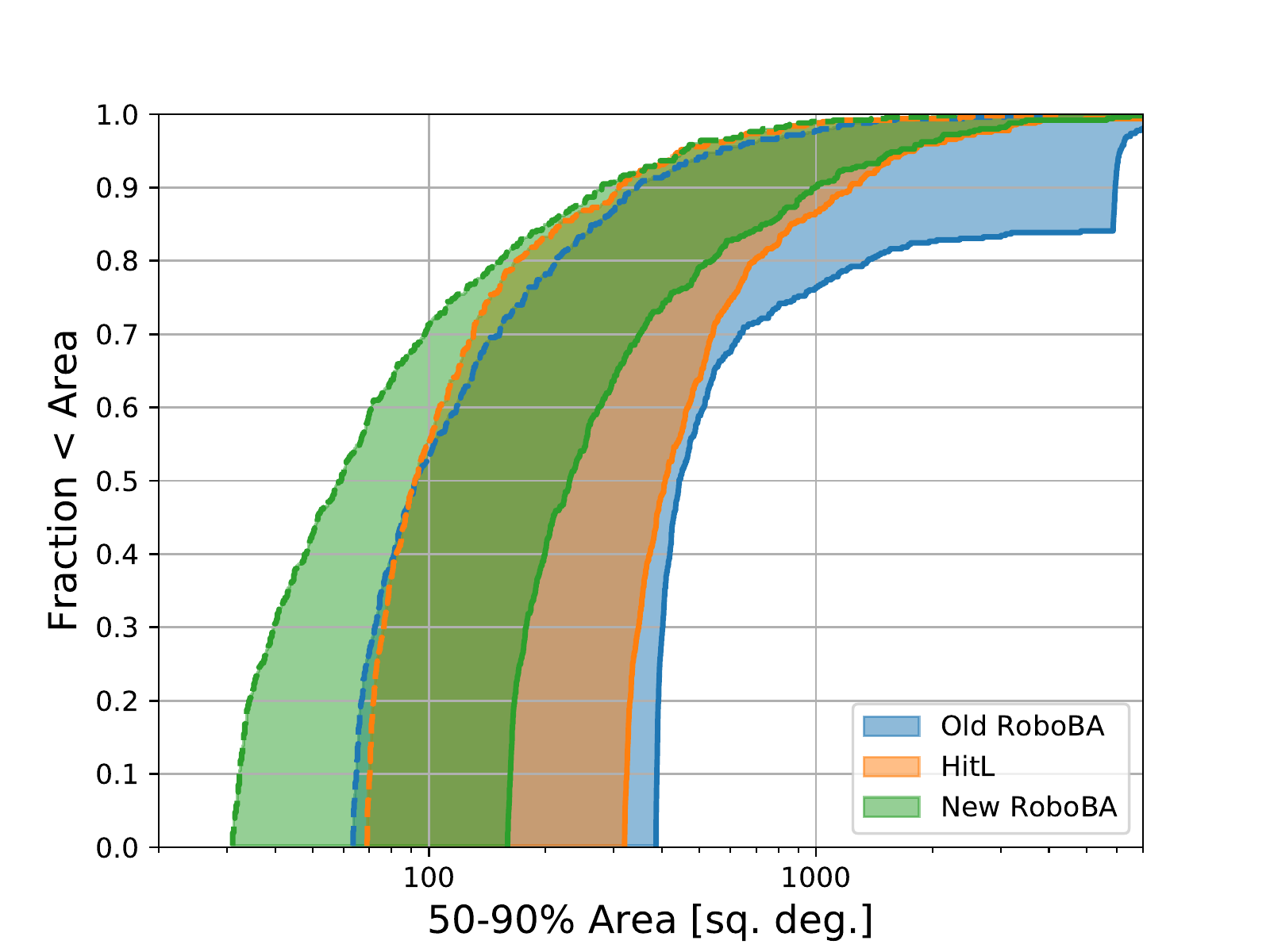}
	\end{center}
\caption{{\it [Left]} The calibration of the \RoboBA\ systematic as evaluated over 505 GRBs, splitting into long and short GRB samples, and splitting into hard, normal, and soft spectra. The distribution with no systematic is shown for comparison.  The purple contours denote the empirical 1, 2, and 3$\sigma$ confidence levels for the calibration. {\it [Right]} Cumulative distributions of the  50\% (dashed line) credible area and 90\% (solid line) credible area for the old and new \RoboBA\ and the HitL localizations.
\label{fig:RoboBA_systematic}}
\end{figure}

For estimating the systematic between \RoboBA-identified long and short GRBs, we divide the sample and fit each separately: 431 long GRBs and 74 short GRBs.  The resulting best-fit systematic models are shown in Table~\ref{tbl:RoboBA_fits}.  Similar to the overall systematic uncertainty estimation, the ratio of GRBs in the core and tail are $\sim $1:1, and the long GRBs overall have a smaller systematic uncertainty compared the short GRBs.  The reason for this could be due to the difference in methodology, or it could be due to the fact that long and short GRBs have different spectra and short GRBs will tend to have a larger systematic if the hard spectrum results in a larger systematic.  Addressing the latter hypothesis, we divide the complete sample based on which spectral template was identified as most appropriate for localization.  This results in 145 GRBs with the hard template, 236 GRBs with the normal template, and 124 with the soft template.  We fit each of these separately and find good fits with the `core+tail' model for each.  Indeed, the hard template has the largest associated systematic, likely due to the fact that the atmospheric scattering component has a larger contribution to the response for harder spectra, thereby making the localization more sensitive to the assumed spectral shape.  All three systematic calibrations are approximately equivalent and are shown in Figure~\ref{fig:RoboBA_systematic}.  As a matter of convenience, we choose to use the long/short systematic model for the new \RoboBA\ throughout the rest of this work as well as in our implementation of the updates, since our \RoboBA\ data pipeline is already configured to handle a short/long systematic model.  We may consider refining this decision in the future to further improve the estimate of the systematic on an individual GRB basis.

While the systematic uncertainty we estimate here is smaller for the new version of the \RoboBA\ compared to the original \RoboBA\ and the HitL localization, a more important metric of localization improvement is the total area of the localization posterior once the systematic component is included.  We show in Figure~\ref{fig:RoboBA_systematic} the comparison of the 50\% and 90\% credible region areas on the sky between the original and updated \RoboBA\ and the HitL.  Note that the HitL localizations are not precisely of the same GRBs, since HitL localizations are rarely performed since 2016, but they were chosen randomly from the pre-2016 set of GRBs. Immediately obvious is the reduction in area on the sky with the new version of the \RoboBA.  The median improvement in the 50\% credible region area is from $\sim 80\ \rm deg^2$ to $\sim 50\ \rm deg^2$, while the median improvement in the 90\% credible region area is from $\sim400\ \rm deg^2$ to $\sim200\ \rm deg^2$. This improvement in the new \RoboBA\ is not only due to resolving the long tail in the systematic, but is a result of an overall improvement in the accuracy of the \RoboBA\ and \dol\ to localize GRBs.

% BALROG-only results: systematics estimation for full sample, bright/weak, long/short, (azimuthal-dependence?)
\subsection{\balrog\ Results}
\begin{figure}
	\begin{center}
		\includegraphics[scale=0.55]{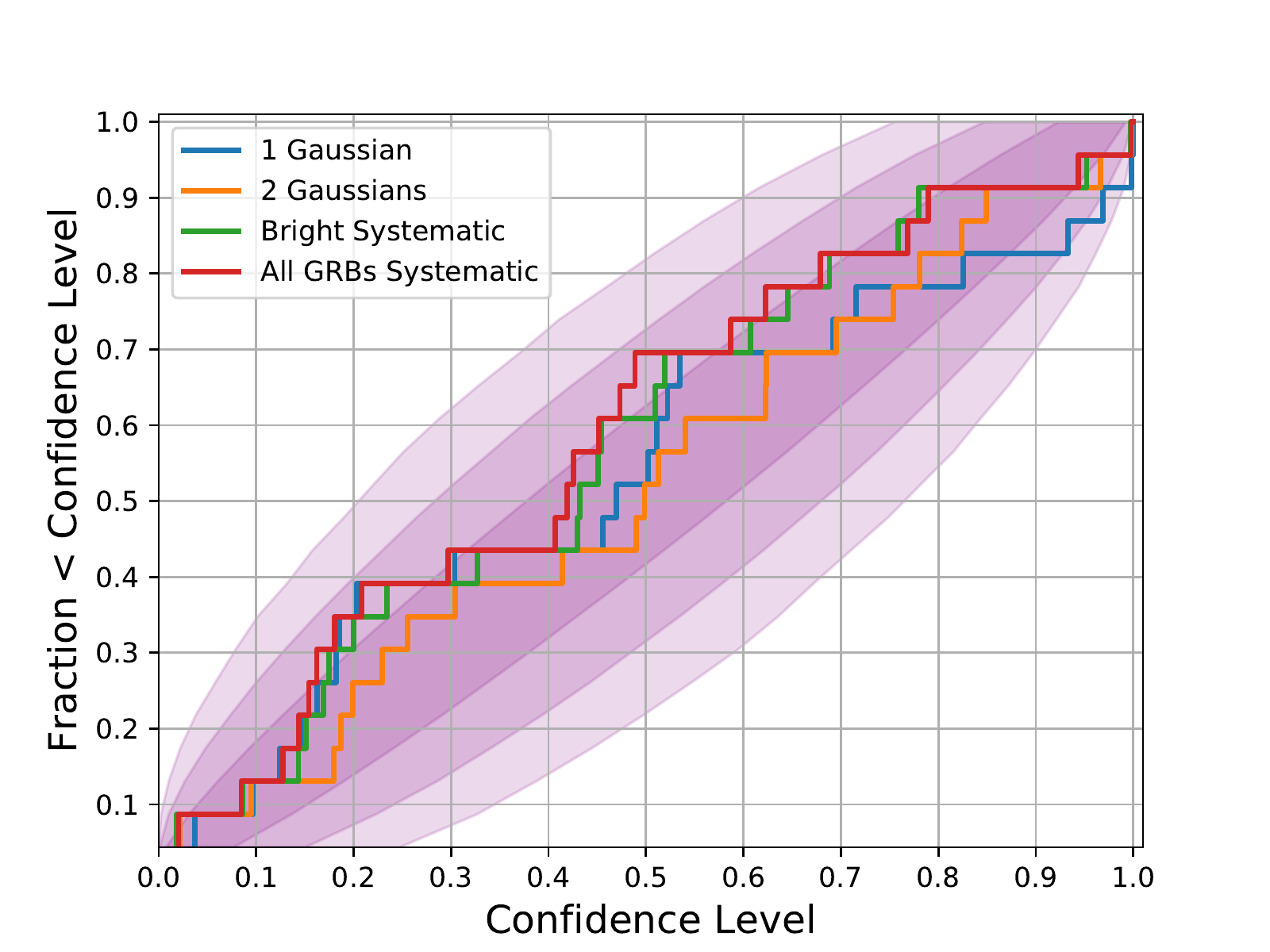}
	\end{center}
\caption{The probability plot for the publicly reported \balrog\ localizations using an estimated systematic models incorporating a single Gaussian and a mixture of two Gaussians.  The two Gaussian `core+tail` model appears to be favored. The systematic models estimated for the bright sample and complete sample of GRBs are consistent with that observed for the publicly reported sample within 2$\sigma$.
\label{fig:realtime_calibration}}
\end{figure}

For \balrog, we can first use this method to estimate the systematic on the public automated localizations performed by \balrog\ since March 2019.  Performing the maximum likelihood estimation, we find that the systematic is likely represented by a mixture of Gaussians, similar to the `core+tail' model discussed earlier.  The core Gaussian has $\sim2.4^\circ$ radius, representing $\sim82\%$ of the localizations and the remaining 18\% of localizations exhibit a $\sim32^\circ$ tail. The fit results, including testing a single Gaussian fit is shown in Table~\ref{tbl:BALROG_fits}, while the calibration including this systematic model is shown in Figure~\ref{fig:realtime_calibration}.

\begin{table}[!h]
\begin{center}
\begin{tabular}{ l | c c c c r }
\hline
\tablecolumns{5}
Sample (\# GRBs) & Model & Core ($^\circ$) & Core \% & Tail ($^\circ$) & Log-Like\\
\hline
\hline
\multirow{2}{*}{Realtime (23)} & Single Gaussian & $3.5\pm0.1$ & -- & -- & --285\\
\cline{2-6} 
                      & Core+Tail & $2.4\pm0.1$ & $82\pm4$ & $32\pm6$ & --267\\
\hline
\multirow{2}{*}{All GRBs (396)} & Single Gaussian & $6.58\pm0.01$ & -- & -- & --87516\\
\cline{2-6} 
                      & Core+Tail & $3.08\pm0.03$ & $74.9\pm0.3$ & $33.3\pm0.6$ & --78911\\
\hline
\multirow{2}{*}{Bright GRBs (230)} & Single Gaussian & $5.60\pm0.02$ & -- & -- & --32051\\
\cline{2-6} 
                      & Core+Tail & $2.68\pm0.03$ & $73.6\pm0.5$ & $32.9\pm1.2$ & --26716\\
\hline
\multirow{2}{*}{Weak GRBs (166)} & Single Gaussian & $8.22\pm 0.09$ & -- & -- & --14246\\
\cline{2-6} 
                      & Core+Tail & $5.44\pm0.29$ & $83.7\pm1.7$ & $37.5\pm0.9$ & --13800\\

\hline
\end{tabular}
\caption{The estimated systematic uncertainty for \balrog\ with different samples
\label{tbl:BALROG_fits}}
\end{center}
\end{table}

\begin{figure}
	\begin{center}
		\includegraphics[scale=0.55]{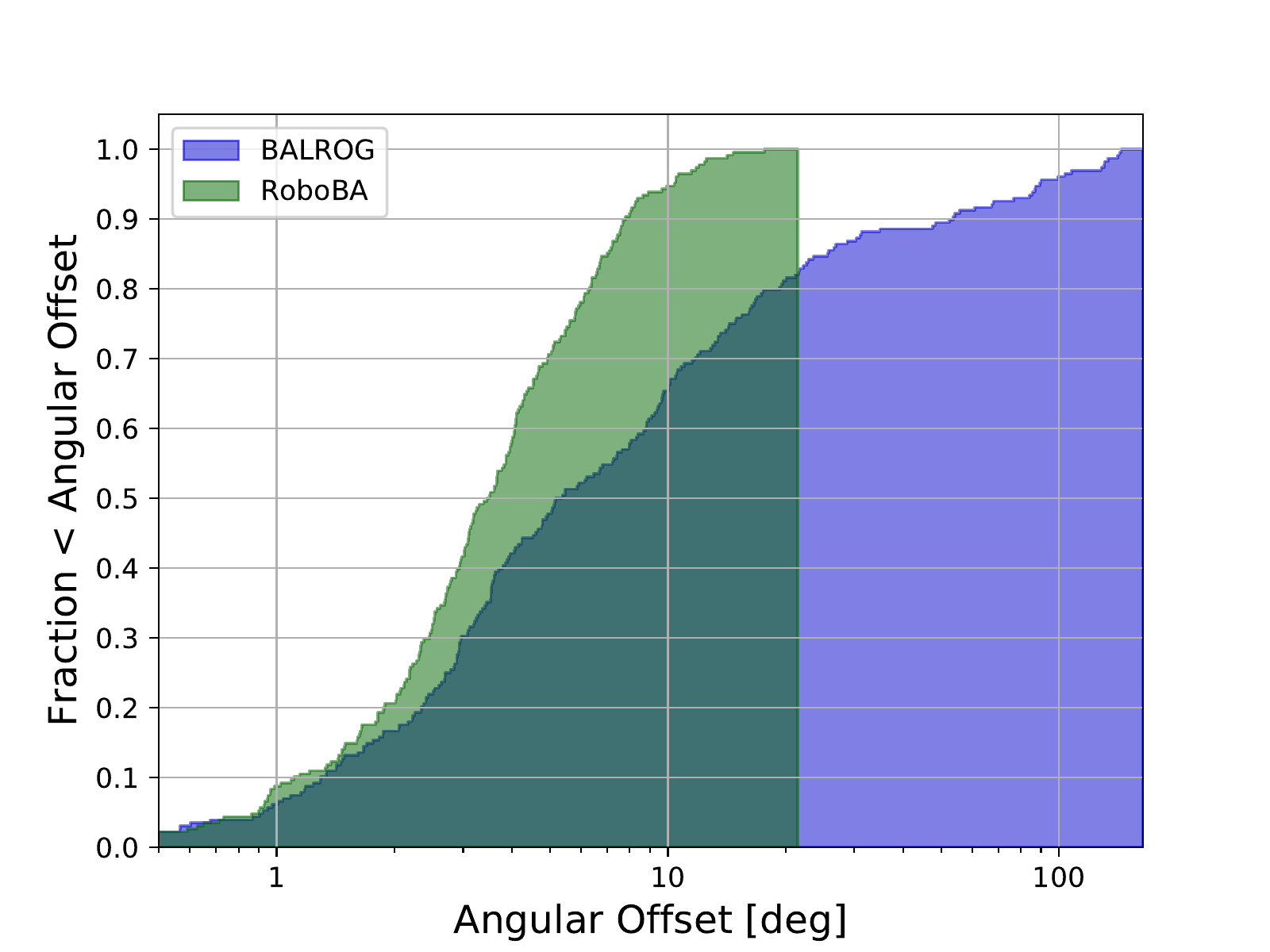}
		\includegraphics[scale=0.55]{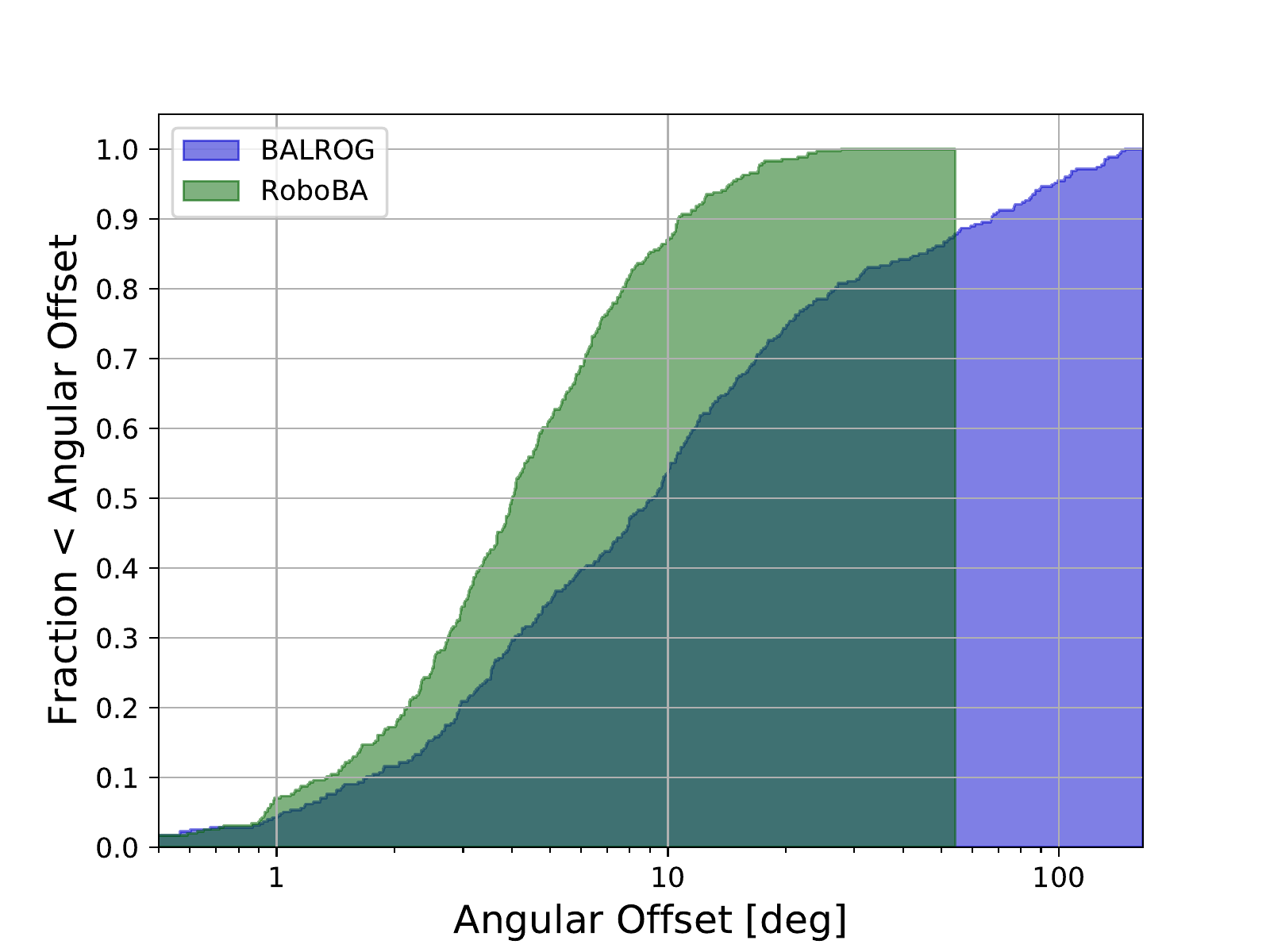}
	\end{center}
\caption{Comparison of the angular offsets from the known location between \RoboBA\ and \balrog\ for the {\it [Left]} bright sample and {\it [Right]} complete sample of GRBs. The long tail in the \balrog\ distribution is from localizations that converged to a completely incorrect part of the sky.
\label{fig:bright_offsets}}
\end{figure}

Similarly, this analysis can be performed on a larger sample of GRBs to provide a better constraint on the systematic and to check consistency with the real-time sample. Since~\citet{Berlato19} defined a `bright' GRB threshold and compiled a sub-sample of GRBs that satisfied that threshold to perform the comparison against the GBM Team's localizations, we use the same definition for the bright sample: 
\begin{equation}
    F_{\rm peak} > (6\ {\rm ph\ cm^{-2}\ s^{-1}}) - (0.857\times10^5\ {\rm erg^{-1}\ s^{-1}})\ S,
\end{equation}
where $F_{\rm peak}$ is the 1-s peak photon flux and $S$ is the energy fluence in 10-1000 keV.  This results in 230 GRBs from the sample described in Section~\ref{sec:sample}. It is important to note that this brightness cut over-samples long GRBs relative to short GRBs as there are only 14 short GRBs in this sample that satisfy the bright criteria.  As a first check of accuracy, we show the head-to-head comparison of the angular offsets of the peak posterior density from the true location between \RoboBA\ and \balrog\ in Figure~\ref{fig:bright_offsets}.  For all but a few percent of the bright GRBs, \RoboBA\ has a small angular offset, with a median offset of $\sim 3^\circ$ and a maximum of $\sim 20^\circ$.  \balrog, however, has a median offset of $\sim 6^\circ$ and a long tail of very large offsets out to $\sim120^\circ$.  Note that this long tail is not a consequence of failed convergence, as those have been removed from the sample; the tail appears to be due to localizations that converged to a completely incorrect part of the sky.

\begin{figure}
	\begin{center}
		\includegraphics[scale=0.35]{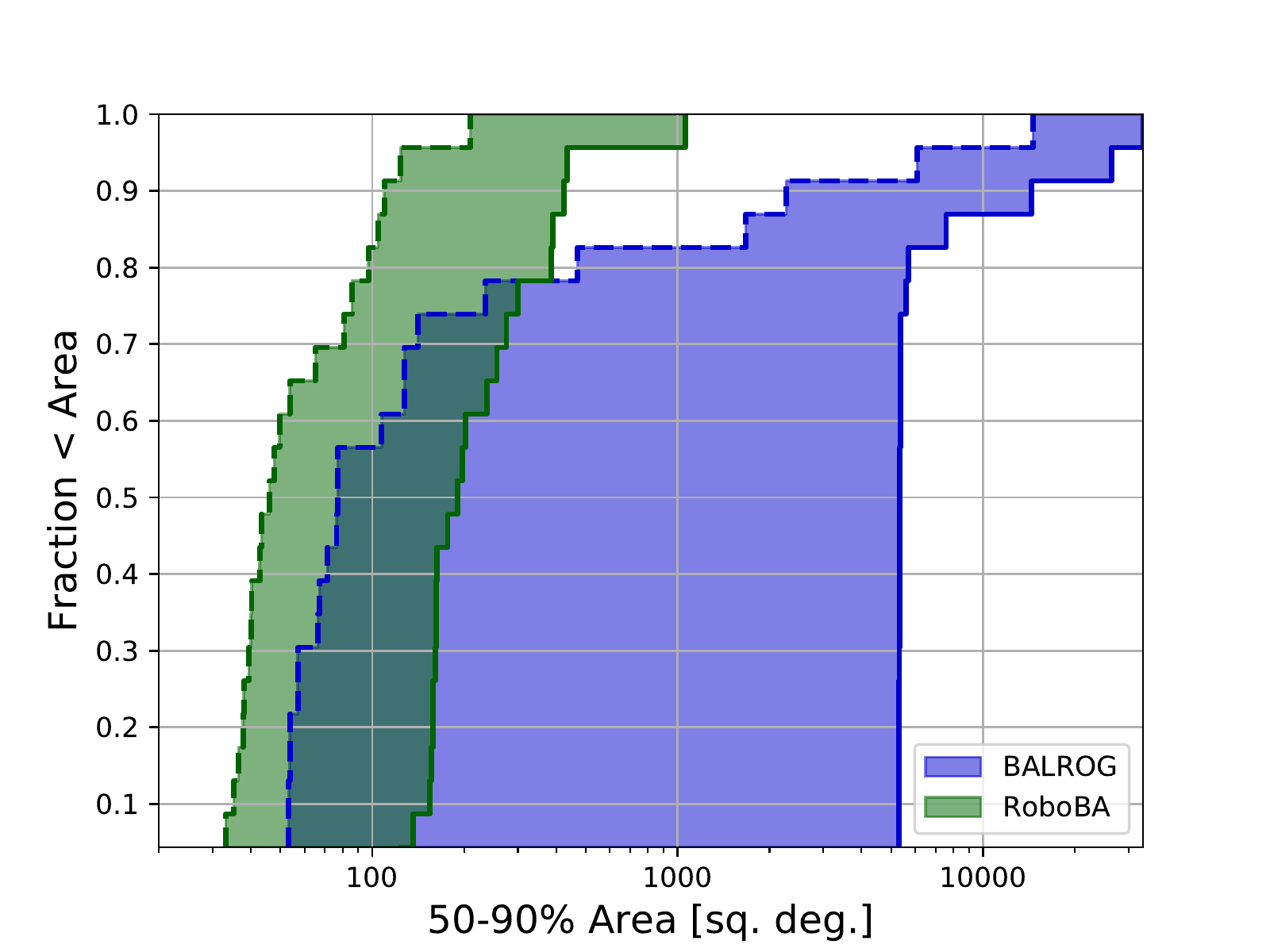}
		\includegraphics[scale=0.35]{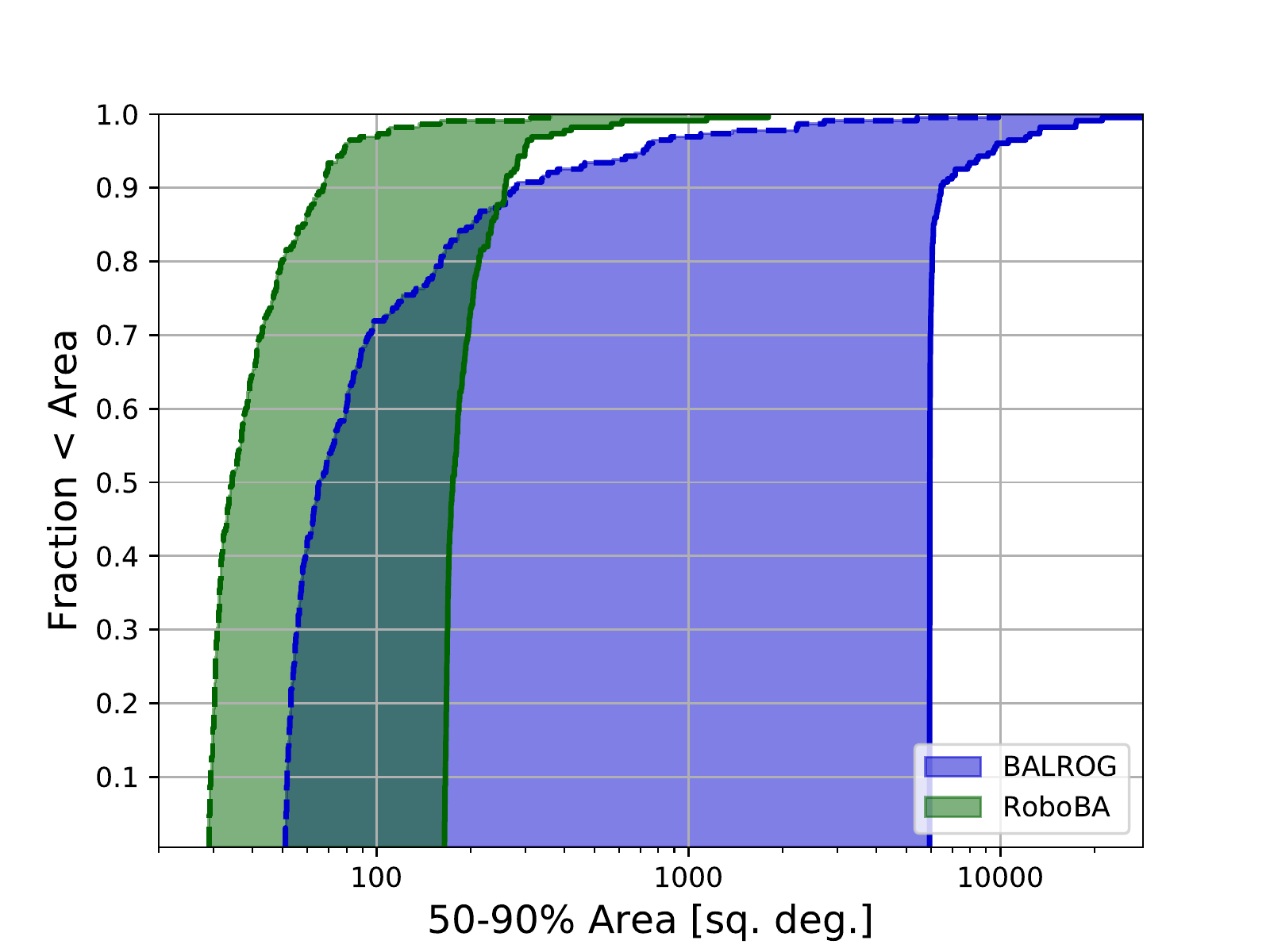}
		\includegraphics[scale=0.35]{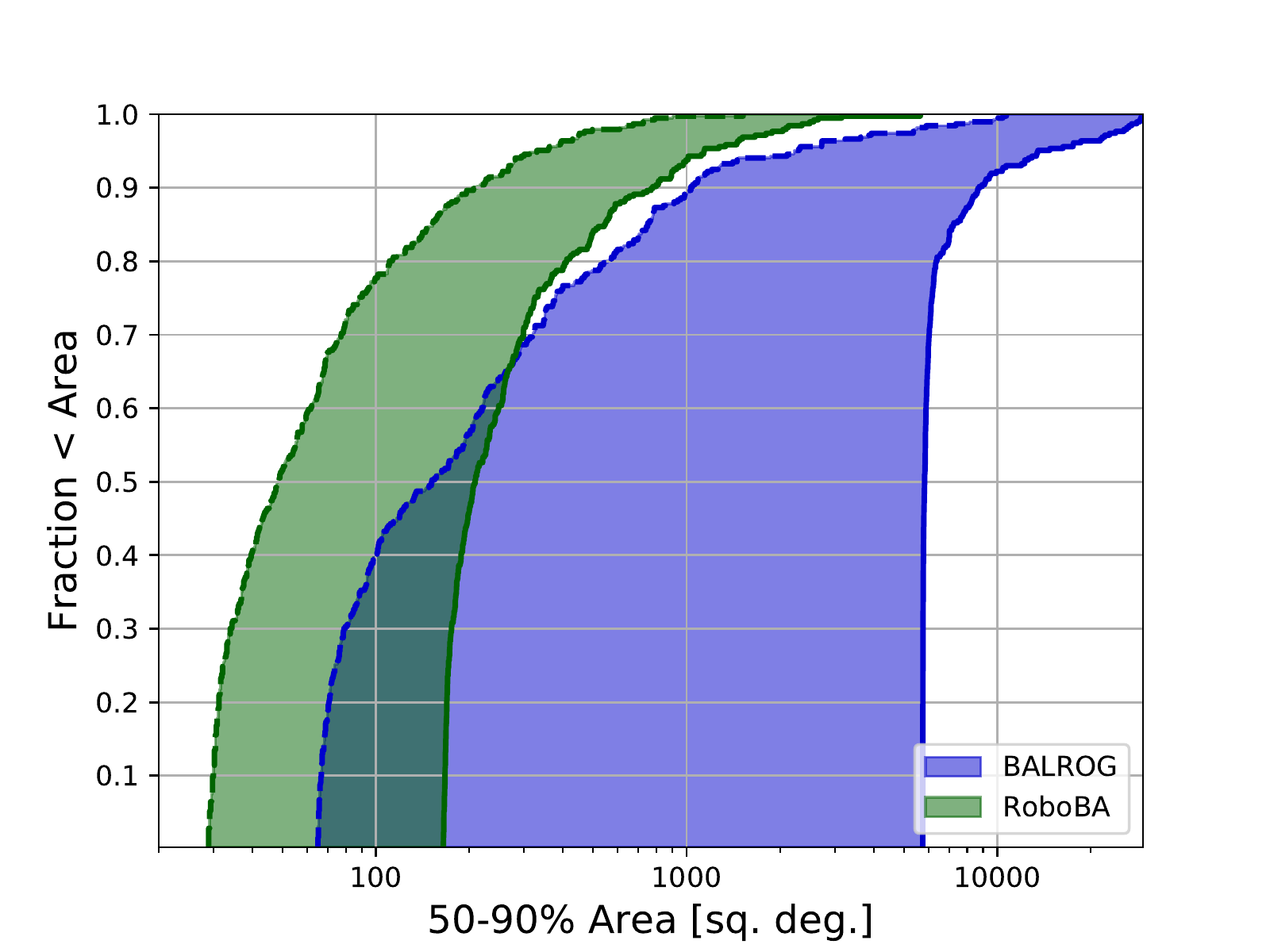}
	\end{center}
\caption{Cumulative distributions of the area for the 50\%--90\% confidence region for the public automated localizations {\it [Left]} for the \RoboBA\ and \balrog, the bright sample {\it [Middle]}, and the full sample {\it [Right]} by applying the corresponding systematics.
\label{fig:bright_area}}
\end{figure}

While the angular offset distribution is consistent with that observed by the real-time localizations shown in Figure~\ref{fig:realtime}, we can also check consistency of the systematic uncertainty for this larger sample.  We fit the `core+tail' model to these GRBs and find that 74\% of the bright GRBs have a $2.7^\circ$ systematic and the remaining 26\% have a $33^\circ$ systematic.  This systematic model is broadly consistent with that found for the real-time localizations, as shown in Figure~\ref{fig:realtime_calibration}. It is not immediately obvious what causes the long tail in the systematic for \balrog, since~\citet{Burgess17} and~\citet{Berlato19} stated that simultaneously fitting the spectrum and location should solve the long tail for previous GBM localizations, however there is strong evidence that this tail is present both in our study as well as the localizations produced in near real-time and distributed publicly.  By applying this more constrained systematic model to the localization posteriors, when can compute the localization area, which is one of the most important figures of merit for GBM localizations. We show in Figure~\ref{fig:bright_area} the comparison of the \RoboBA\ and \balrog\ 50\% and 90\% credible region areas for both the real-time GRBs and the bright sample.

We also estimate the systematic for the full sample of GRBs described in Section~\ref{sec:sample}.  While bright GRBs are considered to be the most affected by systematics, we find that the systematic uncertainty is slightly larger when considering the full sample.  This is likely due to the fact that background modeling and signal selection are increasingly important for {\it weak} GRBs and are of lesser consequence for very bright GRBs, thereby causing different sources of systematic to arise based on the observed brightness of the GRB. This systematic is still broadly consistent with the real-time localizations, but it is a slightly less favorable match than the bright systematic.  The cause of the increased systematic appears to be due to the converse of the bright sample, the `weak' sample of GRBs as the fit parameters show in Table~\ref{tbl:BALROG_fits}.  Indeed, \balrog\ appears to perform systematically {\it worse} for weaker GRBs, possibly a consequence of imprecisely modeling the background, and not carefully minimizing background contamination in the signal selection. Nominally, for statistically-limited signals, the statistical uncertainty should be larger than bright signals, however, it appears that \balrog\ is relatively more overconfident for weaker signals than for bright signals.

\section{Discussion}\label{sec:Discussion}
\begin{figure}[h]
	\begin{center}
		\includegraphics[scale=0.50]{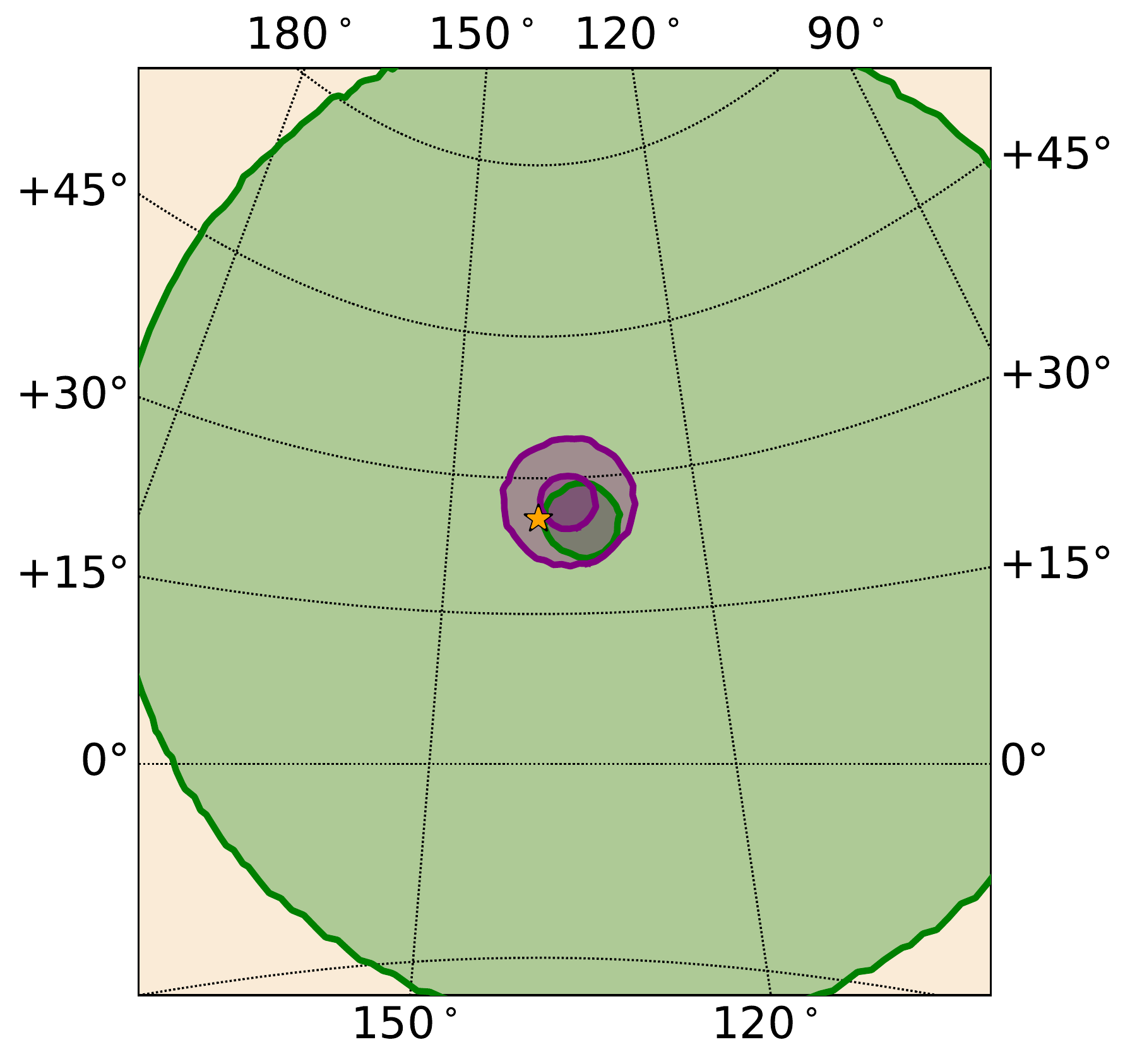}
	\end{center}
\caption{An example comparison of localization posteriors for \RoboBA\ (purple) and \balrog\ (green) once the appropriate systematic models are considered.  The 50\% and 90\% regions are shown, and the true location, determined by the Fermi~LAT, is marked with the gold star.  This localization, of GRB 170522A, is slightly worse than the median angular offset for \RoboBA, and is slightly better than the median angular offset for \balrog.
\label{fig:skymaps}}
\end{figure}

Through a quantitative estimation of the systematic uncertainty for the \RoboBA\ and \balrog\ localizations, we compared the robustness and accuracy of the two algorithms.  Compiled in Table~\ref{tbl:StatSheet} are a number of summary values for the two methods.  The \RoboBA\ is found to be more accurate at localizing GRBs, and is even more accurate with the newest updates detailed in Section~\ref{sec:RoboBA_Improvements}.  Theoretically, it is reasonable to expect that simultaneously estimating the source spectrum and localization would solve some of the systematic uncertainty in GBM localizations, and we find that the different spectral templates the \dol\ uses do indeed produce different systematics.  Clearly if the assumed spectrum is enormously different from the true spectrum, inaccuracies in the localization can be produced, however it is not clear how close the proposed spectrum needs to be to the true spectrum to eliminate this systematic. \citet{Burgess17} discusses the fact that the simultaneous fitting of the spectrum and location can lead to an increase in the area of the statistical uncertainty because it is a more accurate representation of the true uncertainty.  There is evidence that this increase in the statistical uncertainty occurs for most \balrog\ localizations, however, there is still a considerable systematic uncertainty that exists, a larger systematic than even for the localizations produced by the \RoboBA.  In Figure~\ref{fig:skymaps} we show an example of what the localization posteriors look like for an average GRB in the bright sample, once the appropriate systematics are considered.

\begin{table}[h]
\begin{center}
\begin{tabular}{l | l l l}
\hline
 & Original \RoboBA & Updated \RoboBA & \balrog \\
\hline
Systematic Model & long (short)  & long (short) & all (bright)\\
\quad Core (deg.) & 2.6 (3.6) & 1.86 (2.55)  & 3.1 (2.7)\\
\quad Core Fraction (\%) & 65 (98) & 57.9 (39.0) & 74.9 (73.6)\\
\quad Tail (deg.) & 6.0 (29.6) & 4.14 (4.43) & 33.3 (32.9)\\
\hline
Angular Offset & all (bright) & all (bright) & all (bright) \\
\quad Median & $4.3^\circ$ ($3.7^\circ$) & $4.1^\circ$ ($3.5^\circ$) & $9.7^\circ$ ($5.4^\circ$) \\
\quad 90\% Range & $1.3-15^\circ$ ($1.1-11^\circ$) & $1.0-15^\circ$ ($0.9-10^\circ$) &  $1.2-99^\circ$ ($1.0-90^\circ$)\\
\hline
50\% Area (sq. deg.) & all (bright) & all (bright) & all (bright) \\
\quad Median & 83 (69) & 49 (34)  & 151 (66) \\
\quad 90\% Range & 64--461 (64--140) & 29--330 (29-78) & 65--2228 (51--717) \\
90\% Area (sq. deg.) & all (bright) & all (bright) & all (bright) \\
\quad Median & 423 (395) & 209 (175) & 5834 (5936) \\
\quad 90\% Range & 386--5982 (386--5867) & 166--1138 (165--299) & 5756--13507 (5925--9494) \\
\hline
Failure Rate & $\lesssim 15\%$ &  $\lesssim 5\%$ & $\sim 15-20\%$ \\
\hline
CPU Time & $\lesssim 10$ s & $\lesssim 10$ s   & O(hours) \\
\hline
\end{tabular}
\caption{Summary statistics of the old \RoboBA, updated \RoboBA, and \balrog\ localizations. The 90\% range represents the range of values spanning 5\%--95\% of the cumulative distribution.
\label{tbl:StatSheet}}
\end{center}
\end{table}

It is beyond the scope of this work to pinpoint the sources of this additional systematic within \balrog, but it is worth noting the convergence and sensitivity issues that can arise with the \balrog\ algorithm.  For a number of GRBs, the convergence of the localization to the true location is sensitive to the initial conditions, either the initial parameters of Monte Carlo chains or the background fit and signal selection. Indeed, by studying the results from the online version of \balrog, similar issues with convergence appears as well, particularly for weaker and/or short GRBs.   While this can be diagnosed and amended for GRBs with known locations, \balrog\ is only of use for localization if one doesn't know the location {\it a priori}. Our failure rate of $\sim23\%$ in running \balrog\ may be higher than that of the online version ($\sim15-20\%$) even though we followed the guidance and tutorials and attempted to replicate the process of the online algorithm. Potential sources of the differences may lie in our methodology in selecting the detectors to use for each GRB. If that is the case, then the online version of \balrog\ may be utilizing additional information, such as the location provided by the GBM Team as was suggested in~\citet{Berlato19}.  Nevertheless, the fact that our study of \balrog\ localizations finds a systematic consistent with what is being produced by the online version suggests that these failures do not significantly affect our determination of the overall accuracy of \balrog.  Both the automated localizations produced by \balrog\ and our larger study disagree quite strongly with the $1-2^\circ$ systematic quote for \balrog.  This appears to indicate a difference in methodology between the study performed by~\citet{Berlato19} and what is currently implemented for \balrog.  It may be possible to produce a much better localization with \balrog\ if the background, signal selection, and detectors can be tuned for a given GRB, however, that requires considerable input by a human.

Considering that valuable telescope time will be used to follow up GBM GRB localizations, it is important that the community have the best information on the accuracy of the available algorithms. Our results show that the GBM Team's current localizations, primarily those from \RoboBA, are more accurate on average than \balrog, require a smaller systematic uncertainty, have a considerably smaller area covered by the localization posterior, are more robust against failure, and are more computationally efficient leading to a smaller reporting latency. Furthermore, given that the systematic uncertainty provided publicly for \balrog\ considerably underestimates the uncertainty achieved, there is risk of follow-up observations tiling incorrect regions of the sky.
Similarly, spatially correlative studies are at risk of finding false correlations, or missing real correlations, due to the underestimated systematic. Implementation of improvements to \RoboBA\ have significantly improved its accuracy and have reduced the sky area for follow-up searches, and we will continue to investigate additional improvements to the \RoboBA\ and \dol\ algorithms.

\section*{Acknowledgments}
We thank  F.~Berlato and the authors of \balrog\ for providing assistance, guidance, and tutorials in getting \balrog\ installed and operating. This work made use of Astropy, Healpy, SciPy, and the Multi-Mission Maximum Likelihood framework~\citep[3ML;][]{Vianello15}. The USRA co-authors gratefully acknowledge NASA funding through contract NNM13AA43C. The UAH co-authors gratefully acknowledge NASA funding from co-operative agreement NNM11AA01A and that this work was made possible in part by a grant of high performance computing resources and technical support from the Alabama Supercomputer Authority. E.~B. is supported by an appointment to the NASA Postdoctoral Program at the Goddard Space Flight Center, and C.~M. is supported by an appointment to the NASA Postdoctoral Program at the Marshall Space Flight Center, administered by Universities Space Research Association under contract with NASA. C.~M.~H., D.~K. and C.~A.~W.-H. gratefully acknowledge NASA funding through the Fermi GBM project. Support for the German contribution to GBM was provided by the Bundesministerium f{\"u}r Bildung und Forschung (BMBF) via the
Deutsches Zentrum f{\"u}r Luft und Raumfahrt (DLR) under contract number 50 QV 0301. 
%A.v.K. was supported by the Bundesministeriums fÃŒr Wirtschaft und Technologie (BMWi) through DLR grant 50 OG 1101. 

\section{Appendix}\label{sec:Appendix}
The following figures are sky maps comparing the localizations from the original \RoboBA\ and \balrog\ algorithms for each of the 23 GRBs contained in Table~\ref{tbl:realtime}.  The \balrog\ localizations shown here are from the \balrog\ webpage reporting their near-realtime results, and we incorporate the prescribed systematic as determined on the corresponding webpage or GCN notice. The 50\% and 90\% confidence regions are shown for both algorithms, the GBM \RoboBA\ localizations are represented by the purple filled contours, and the \balrog\ localizations are represented by green contours.  The known location of each GRB is marked by a black star.  As can be seen, 9--10 of the known locations exist outside the \balrog\ 90\% confidence region, while only one of the known locations exists outside the \RoboBA\ 90\% confidence region (GRB 190606A).  Additionally, it appears there are 2 non-convergent localizations produced by \balrog: GRBs 190515A and 190606A, and another poorly-converged localization: GRB 190828B.

\begin{figure}[h]
	\begin{center}
		\includegraphics[width=7.5cm]{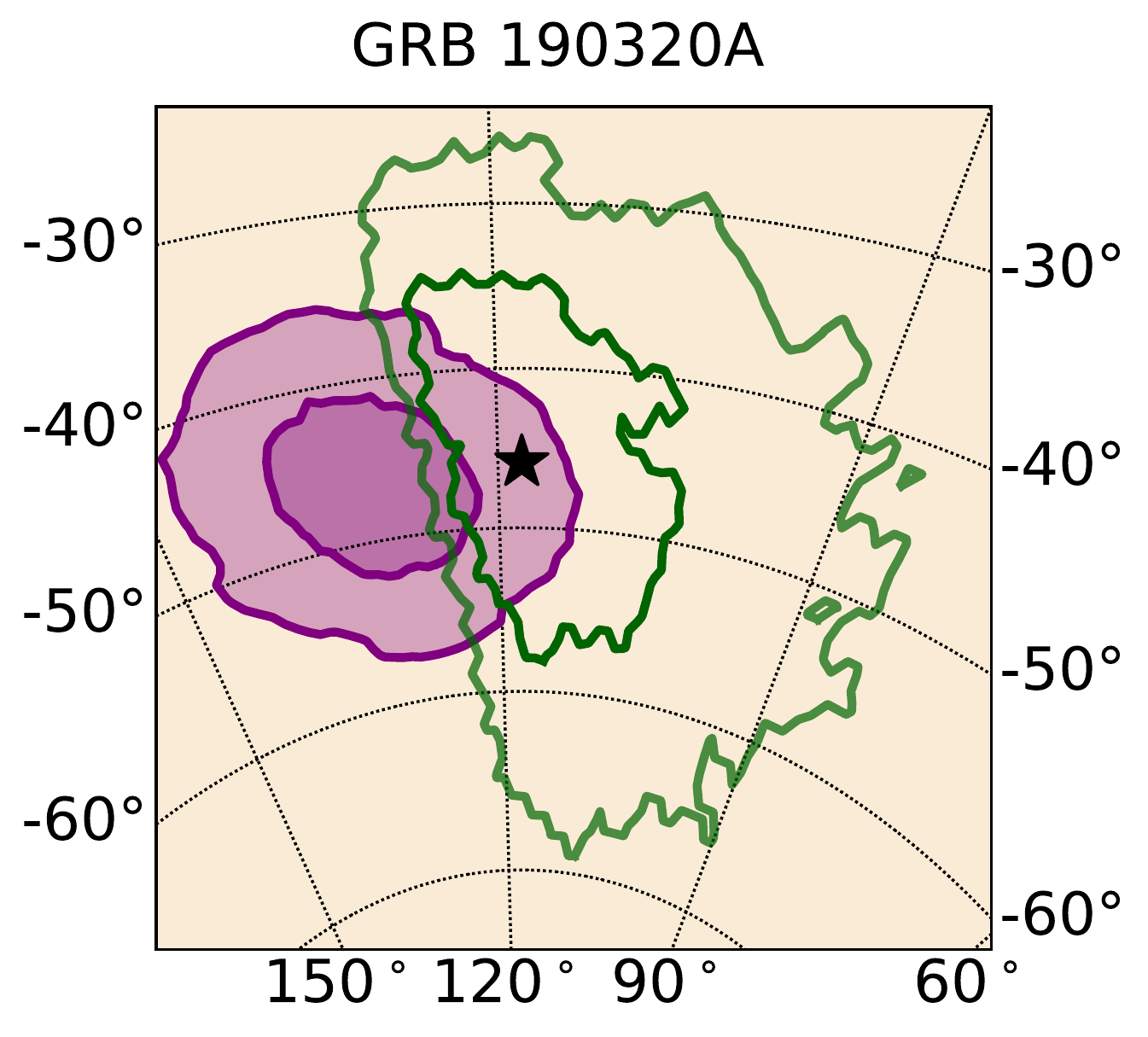}
		\includegraphics[width=7.5cm]{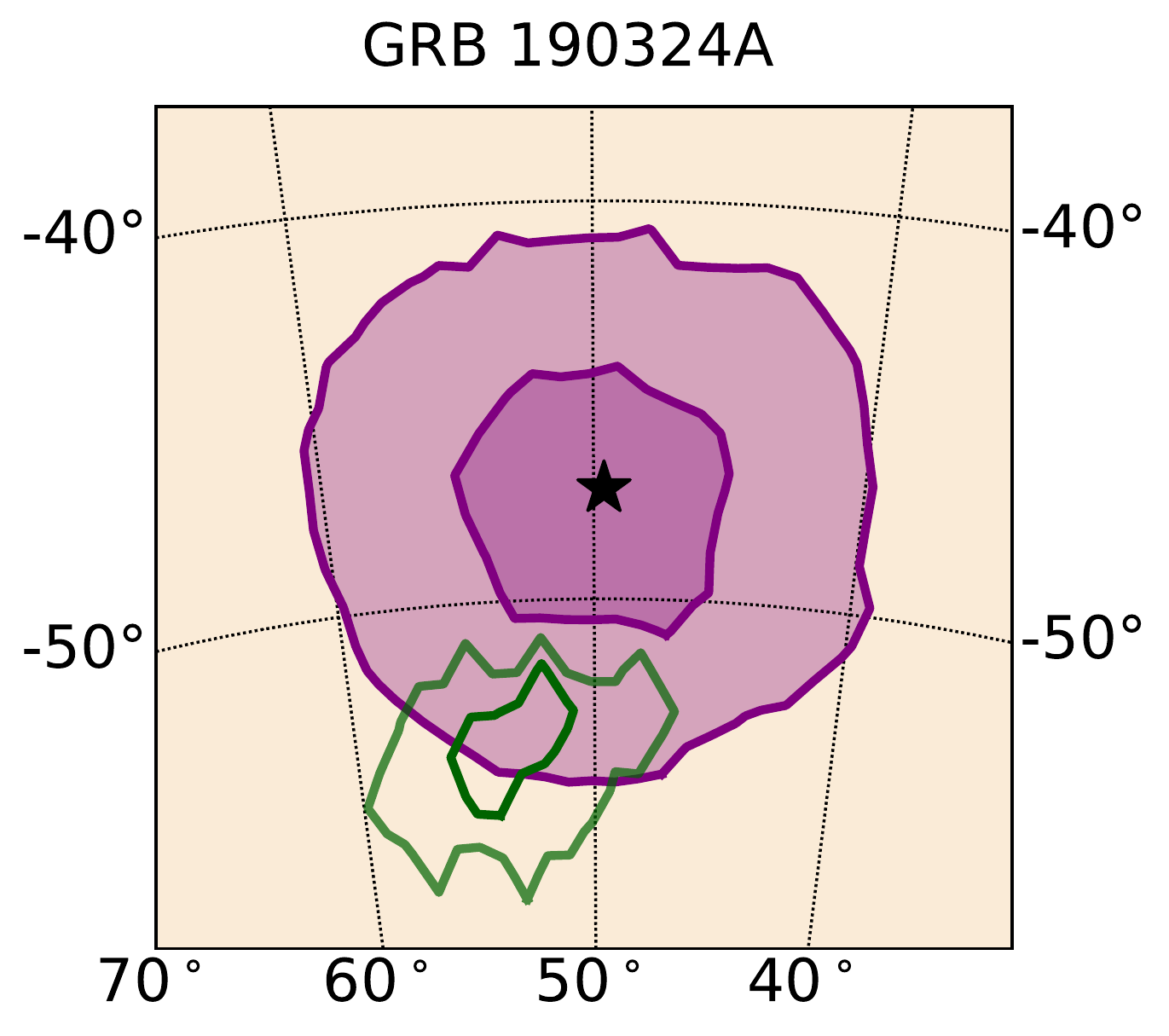}\\
		\includegraphics[width=7.5cm]{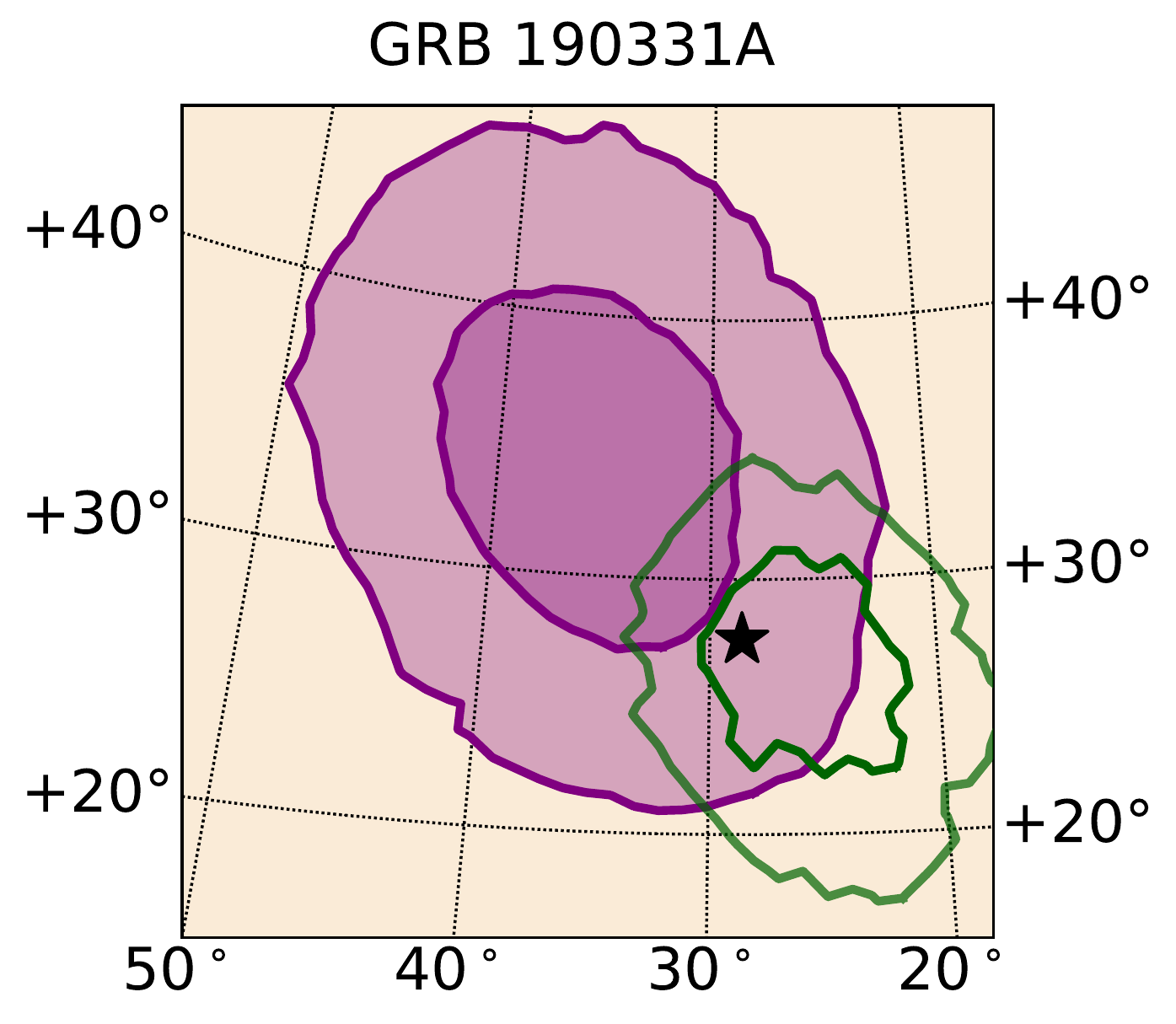}
		\includegraphics[width=7.5cm]{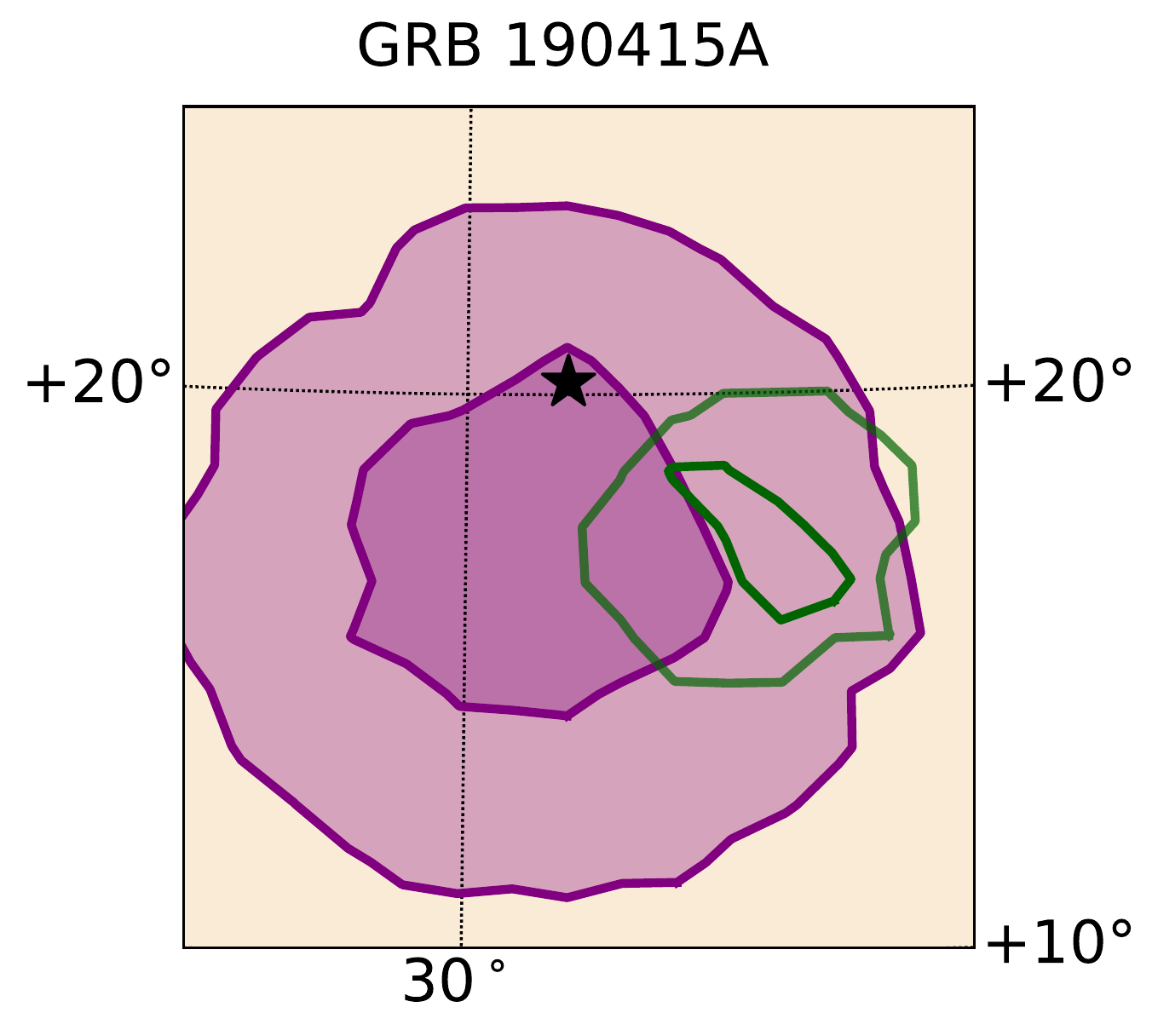}\\
		\includegraphics[width=7.5cm]{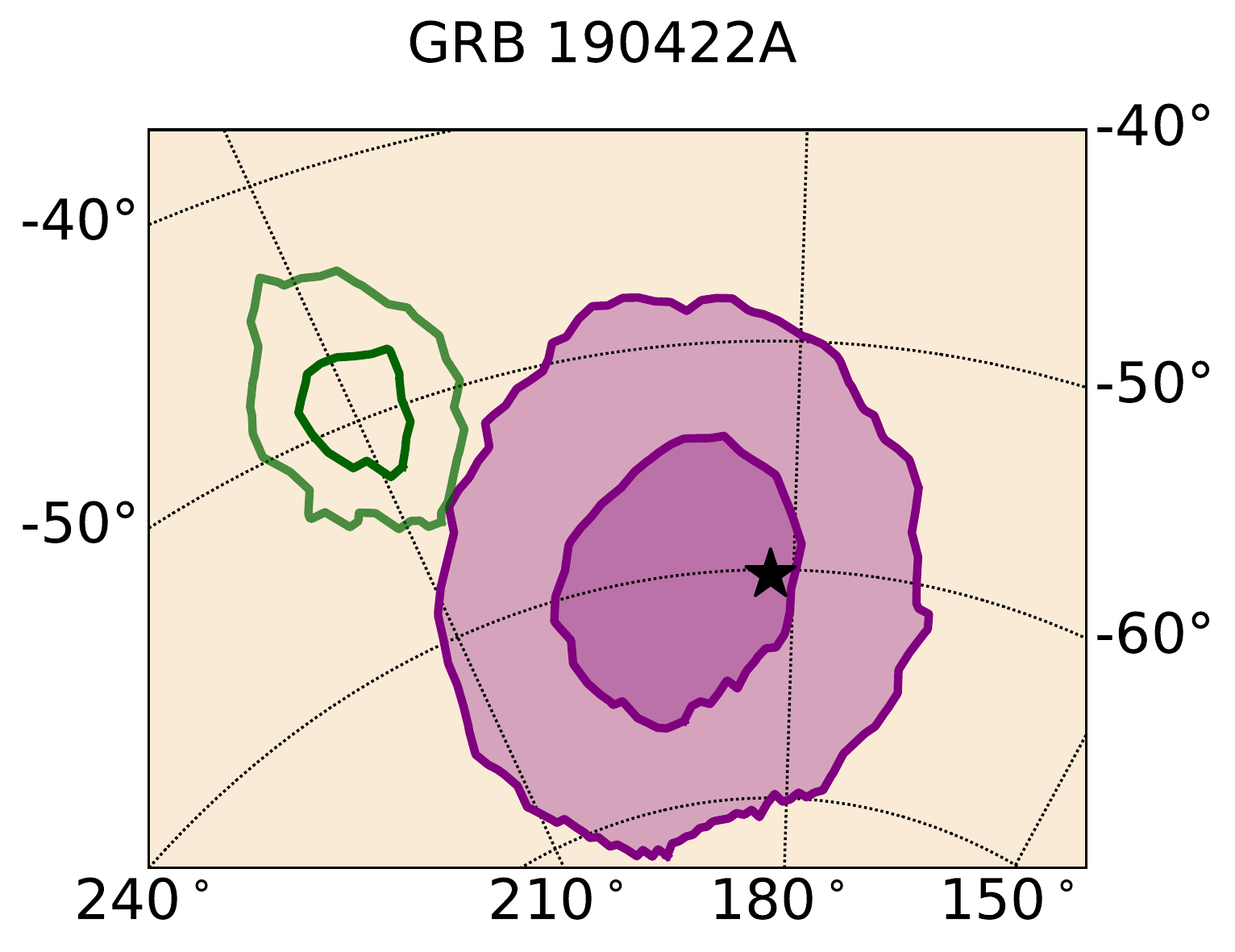}	
		\includegraphics[width=7.5cm]{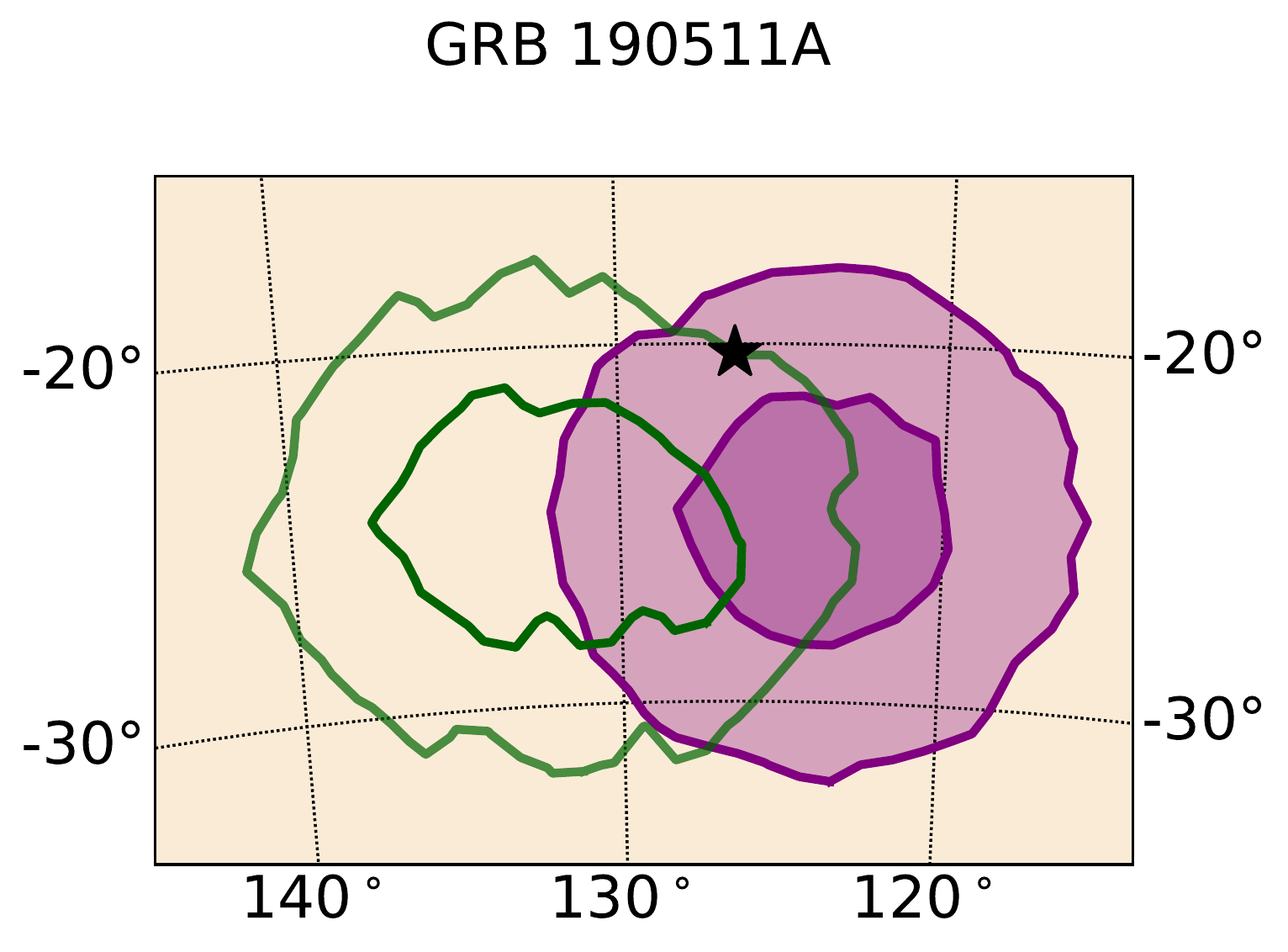}\\		
	\end{center}
\end{figure}

\begin{figure}[h]
	\begin{center}
		\includegraphics[width=15cm]{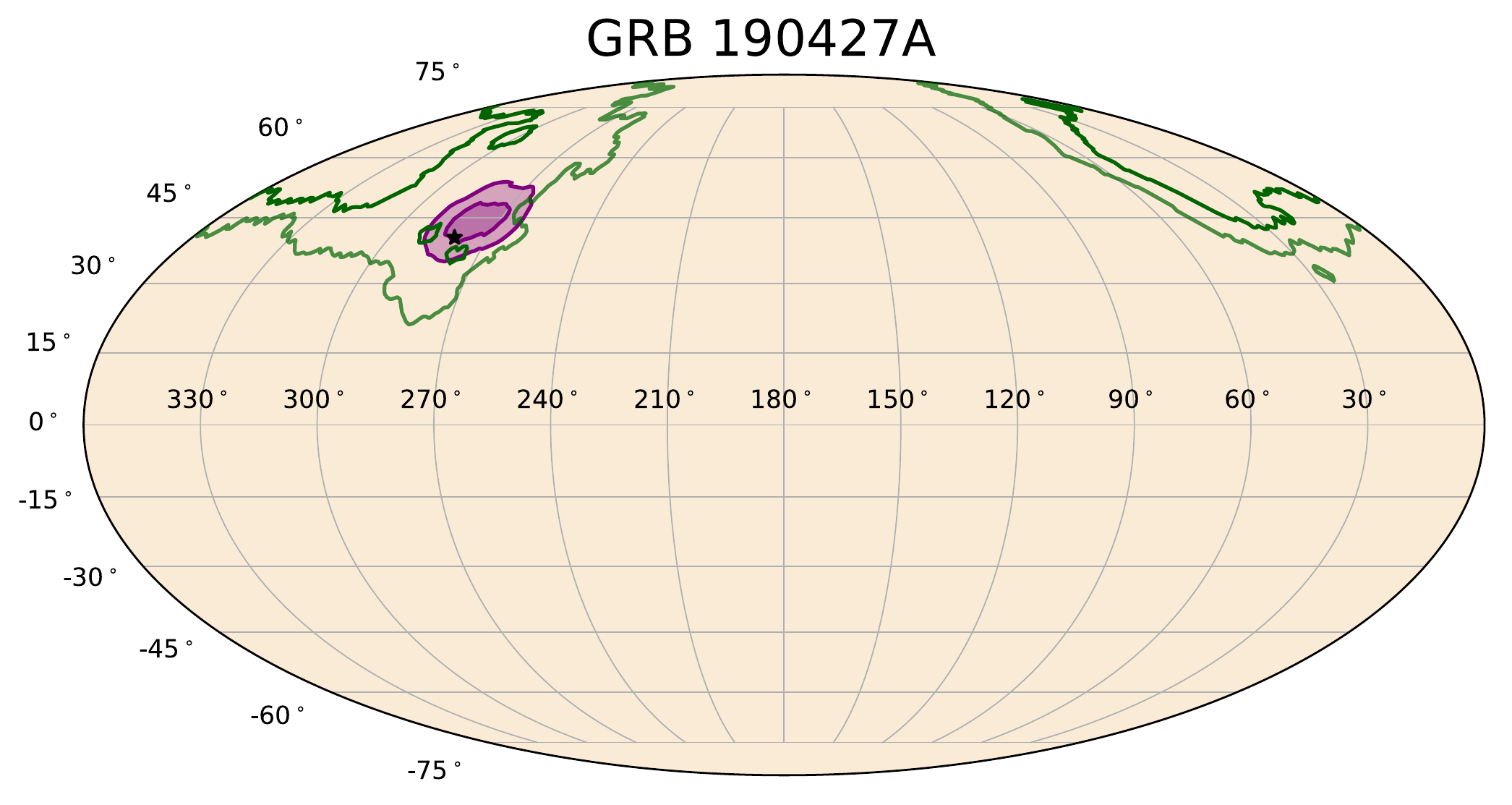}\\		
		\includegraphics[width=15cm]{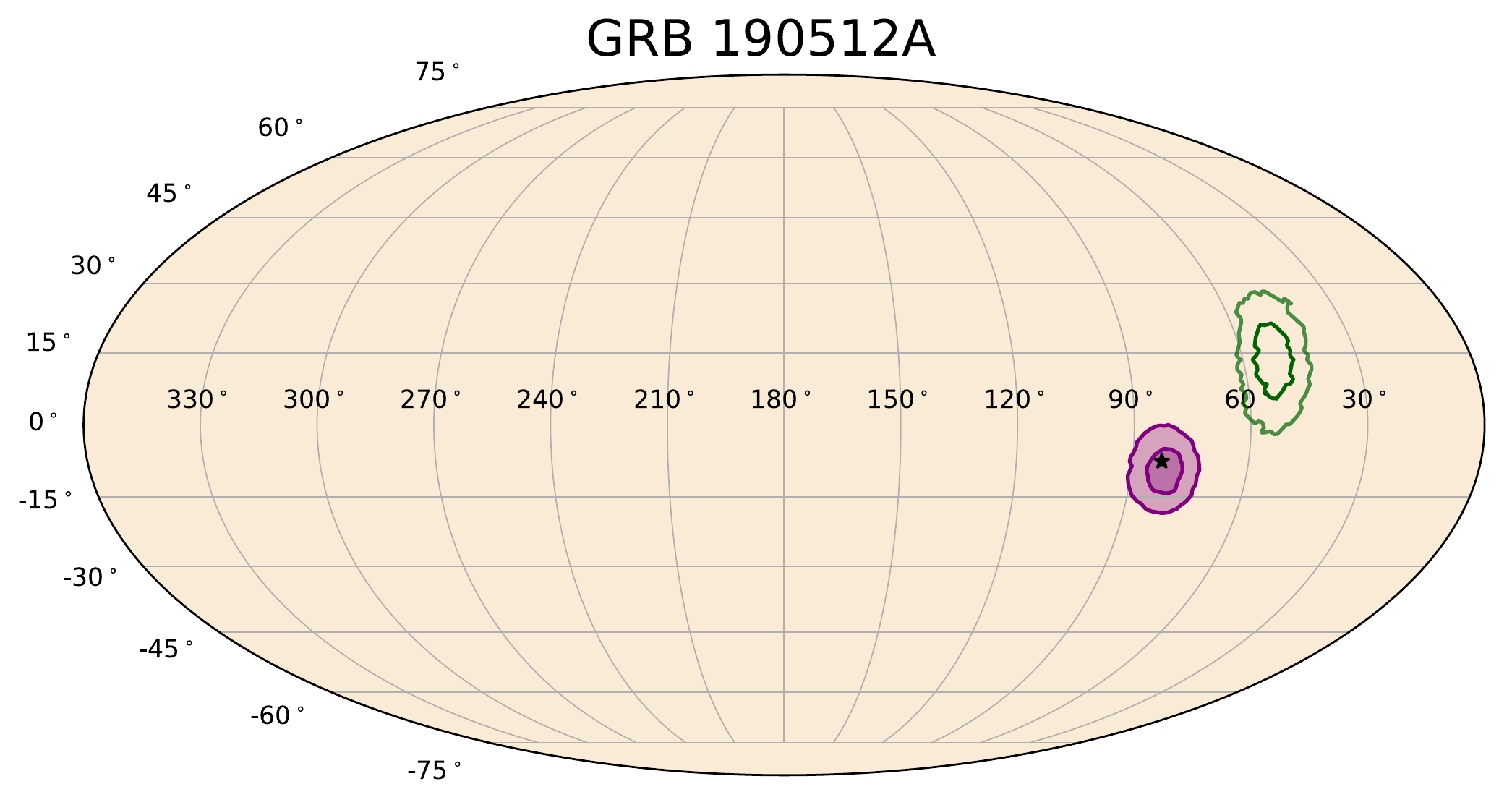}\\
		\includegraphics[width=15cm]{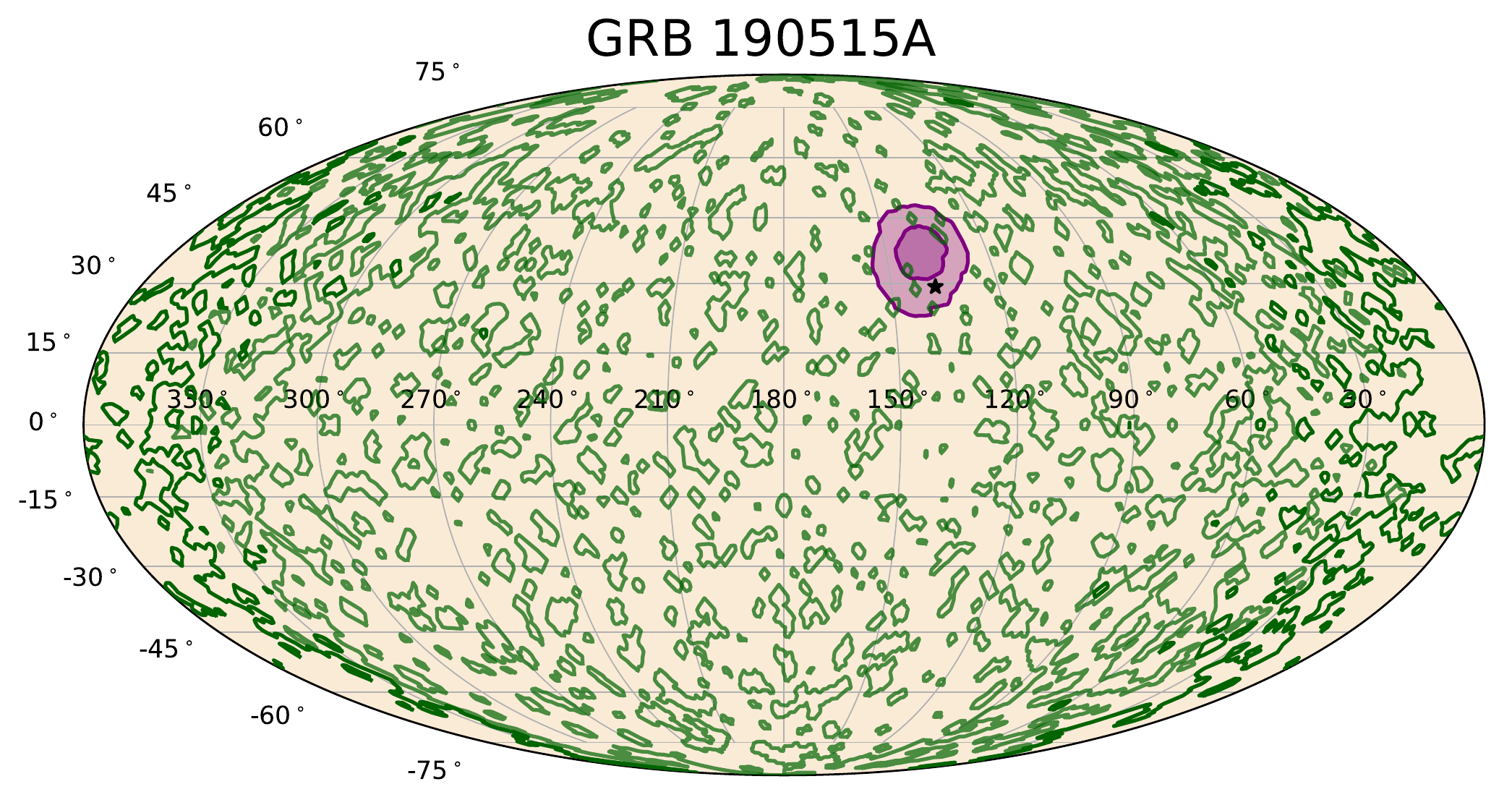}		
	\end{center}
\end{figure}

\begin{figure}[h]
	\begin{center}
		\includegraphics[width=7.5cm]{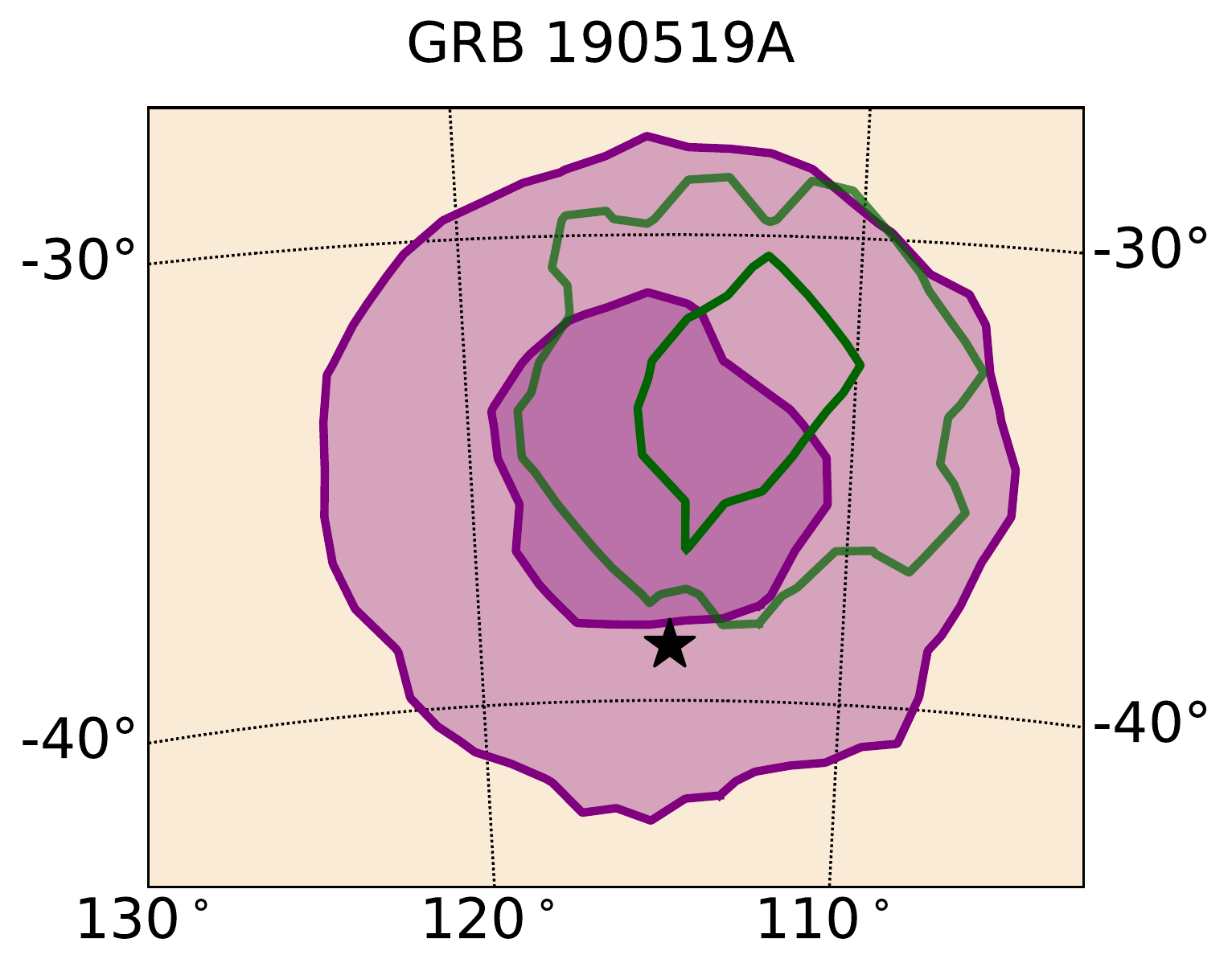}	
		\includegraphics[width=7.5cm]{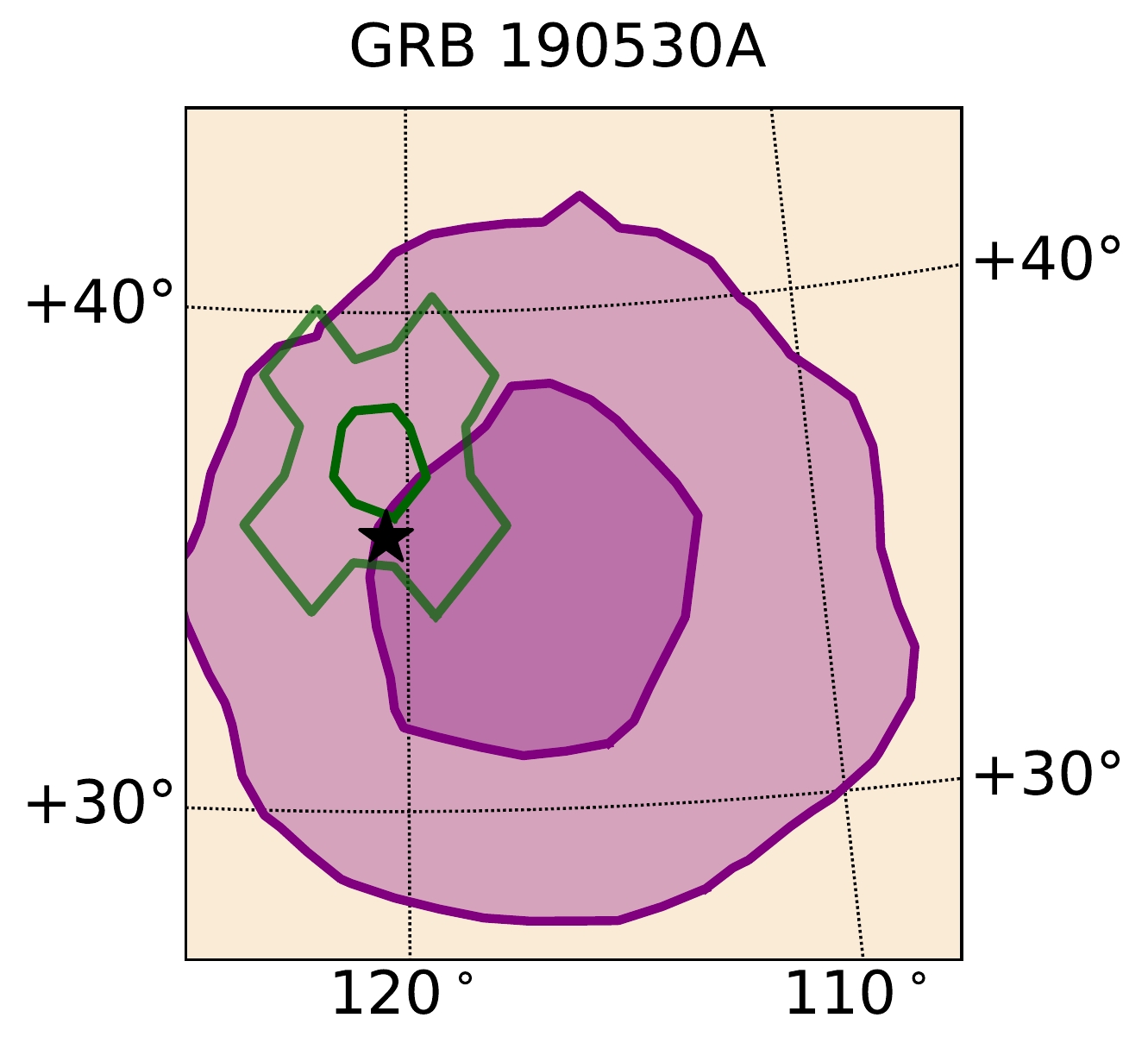}\\		
		\includegraphics[width=7.5cm]{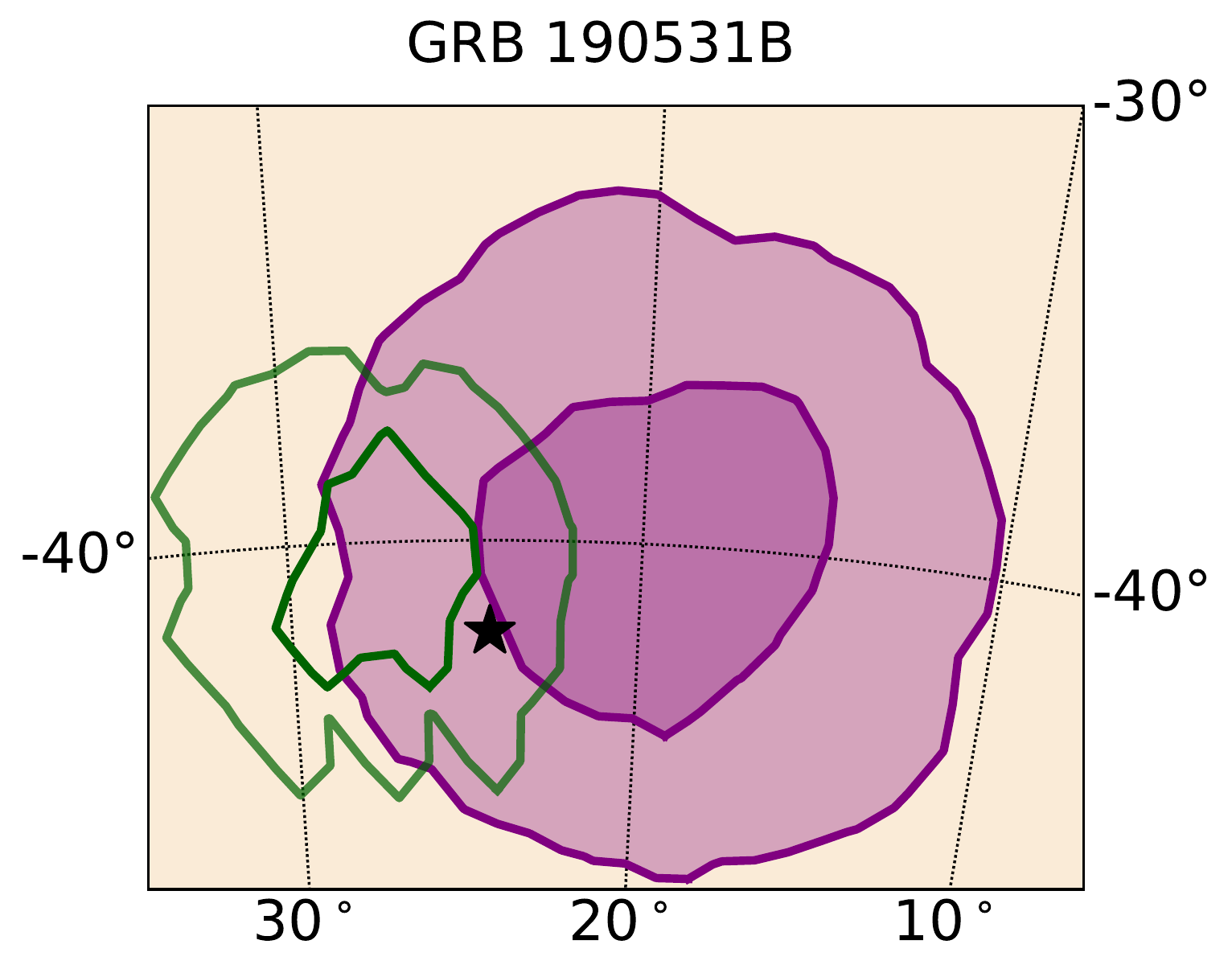}\\	
		\includegraphics[width=15cm]{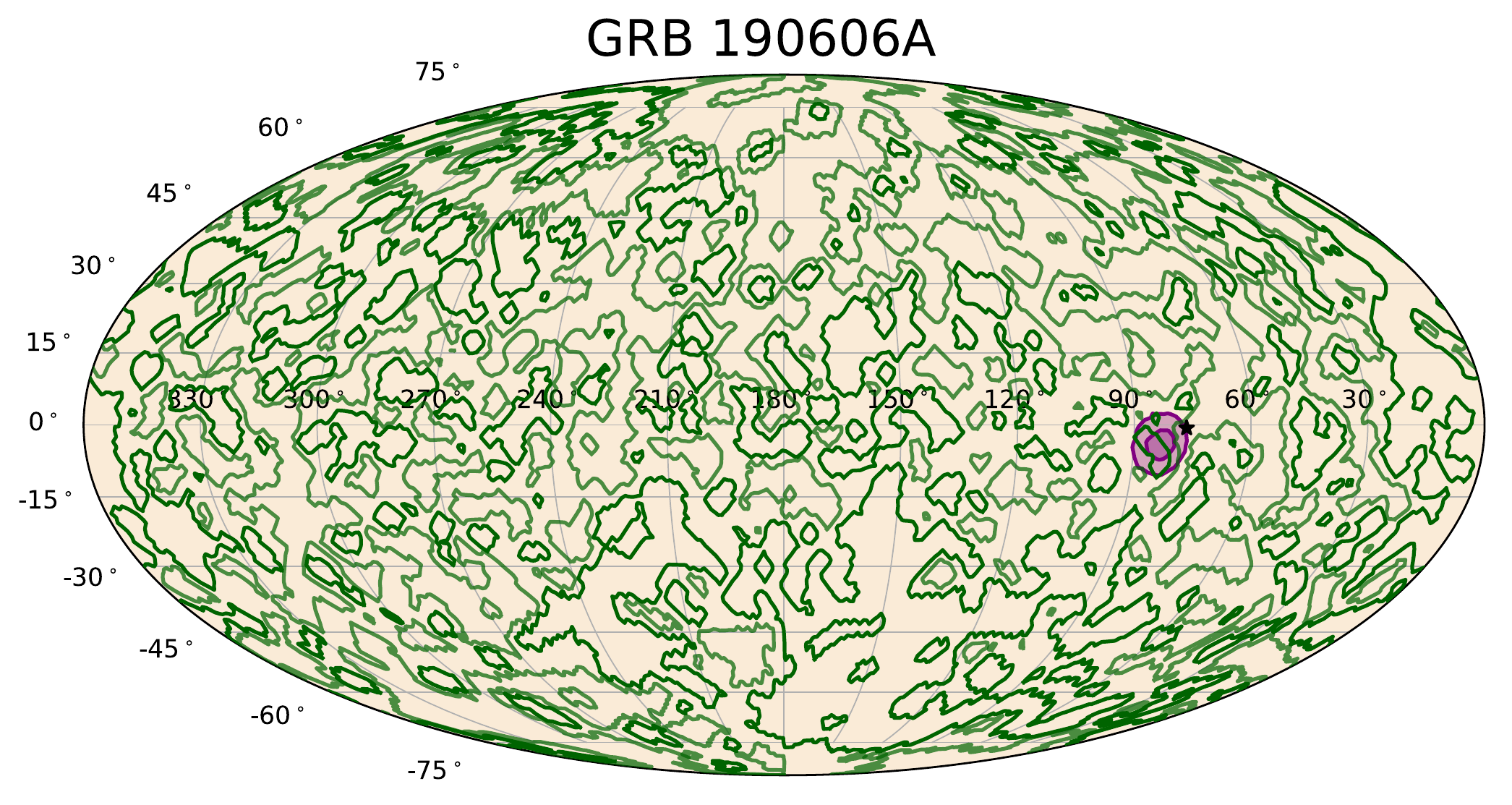}		
	\end{center}
\end{figure}

\begin{figure}[h]
	\begin{center}
		\includegraphics[width=7.5cm]{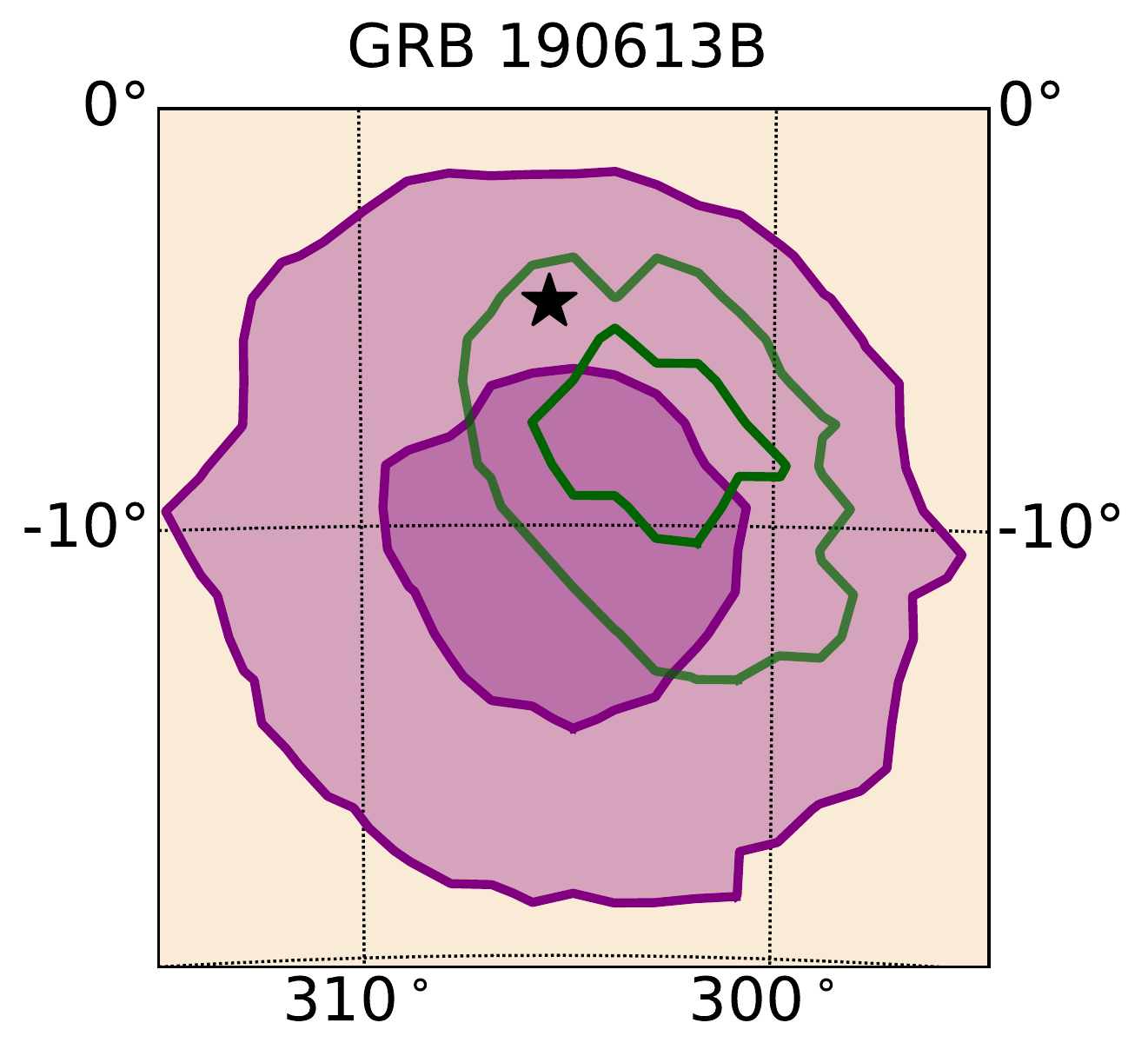}	
		\includegraphics[width=7.5cm]{190530430.pdf}\\		
		\includegraphics[width=7.5cm]{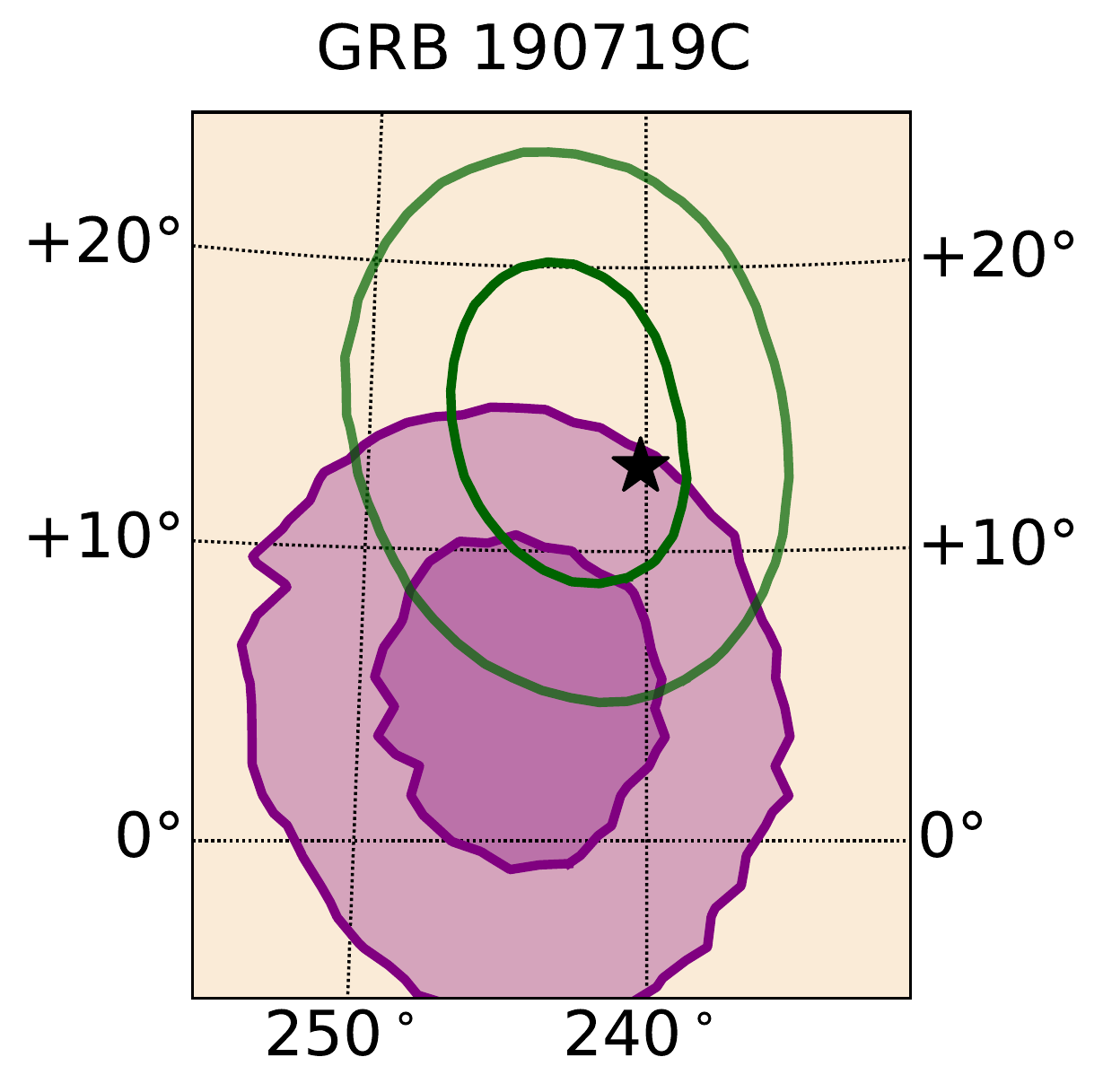}		
		\includegraphics[width=7.5cm]{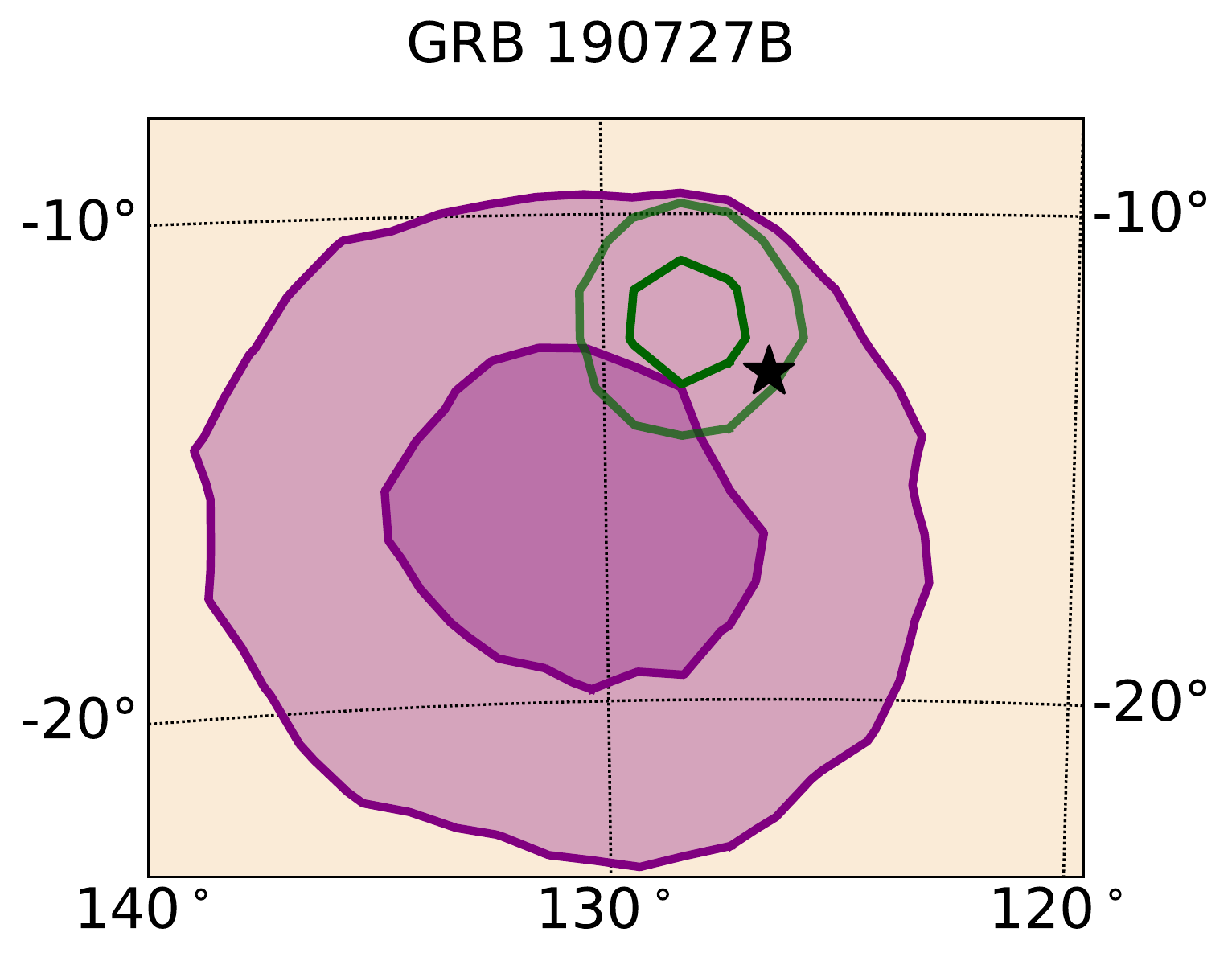}\\
		\includegraphics[width=7.5cm]{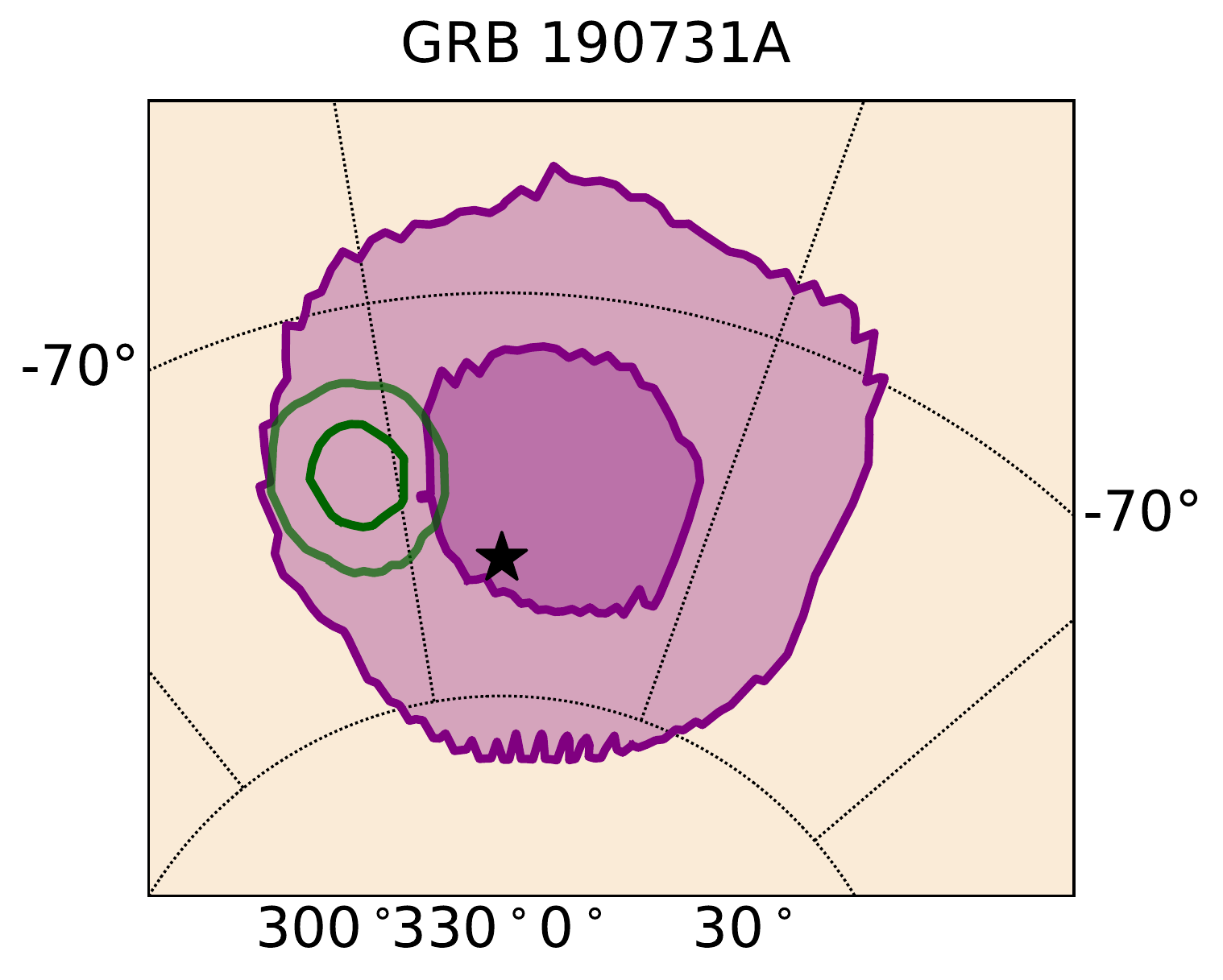}	
		\includegraphics[width=7.5cm]{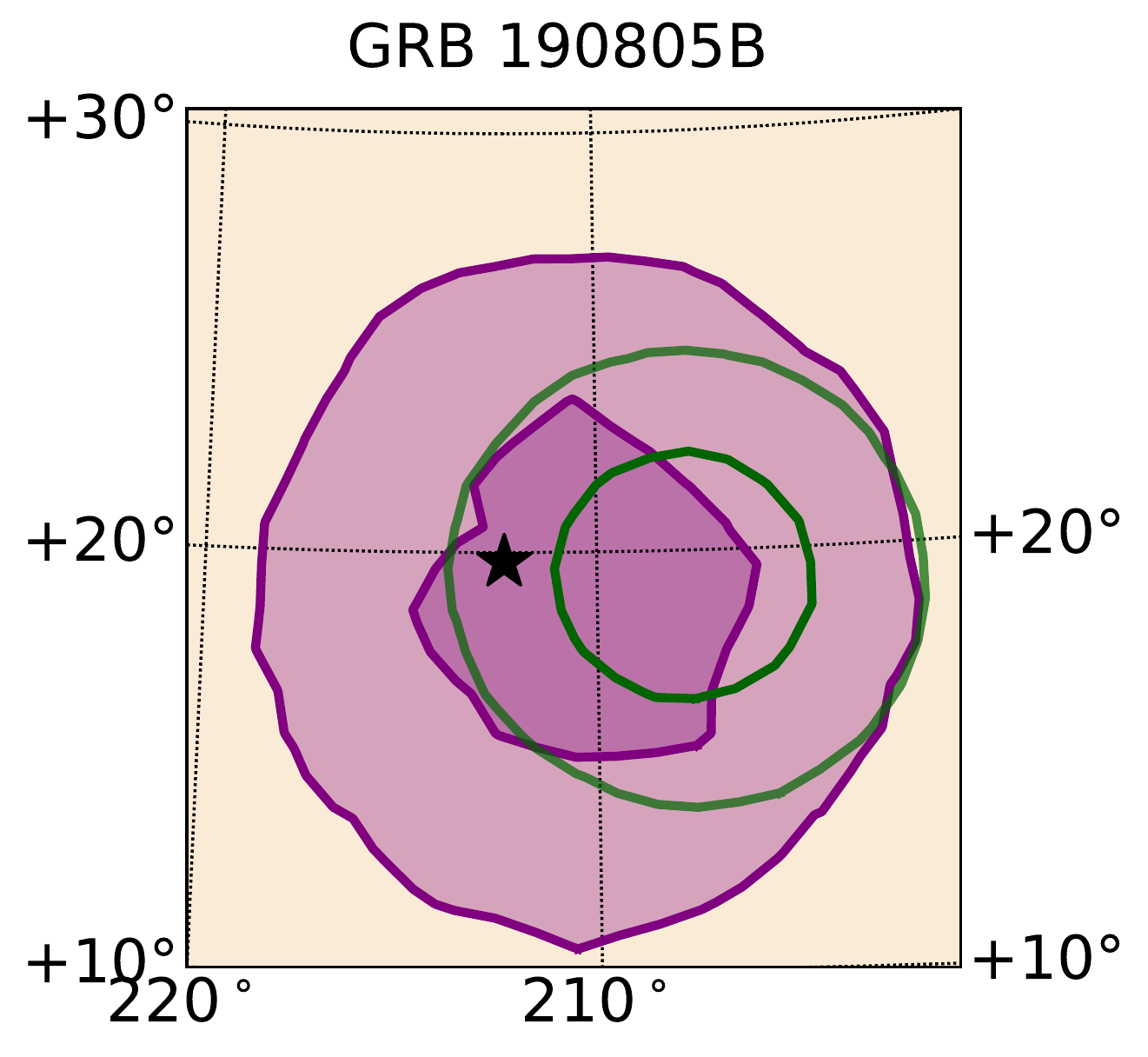}
	\end{center}
\end{figure}

\begin{figure}[h]
	\begin{center}
		\includegraphics[width=7.5cm]{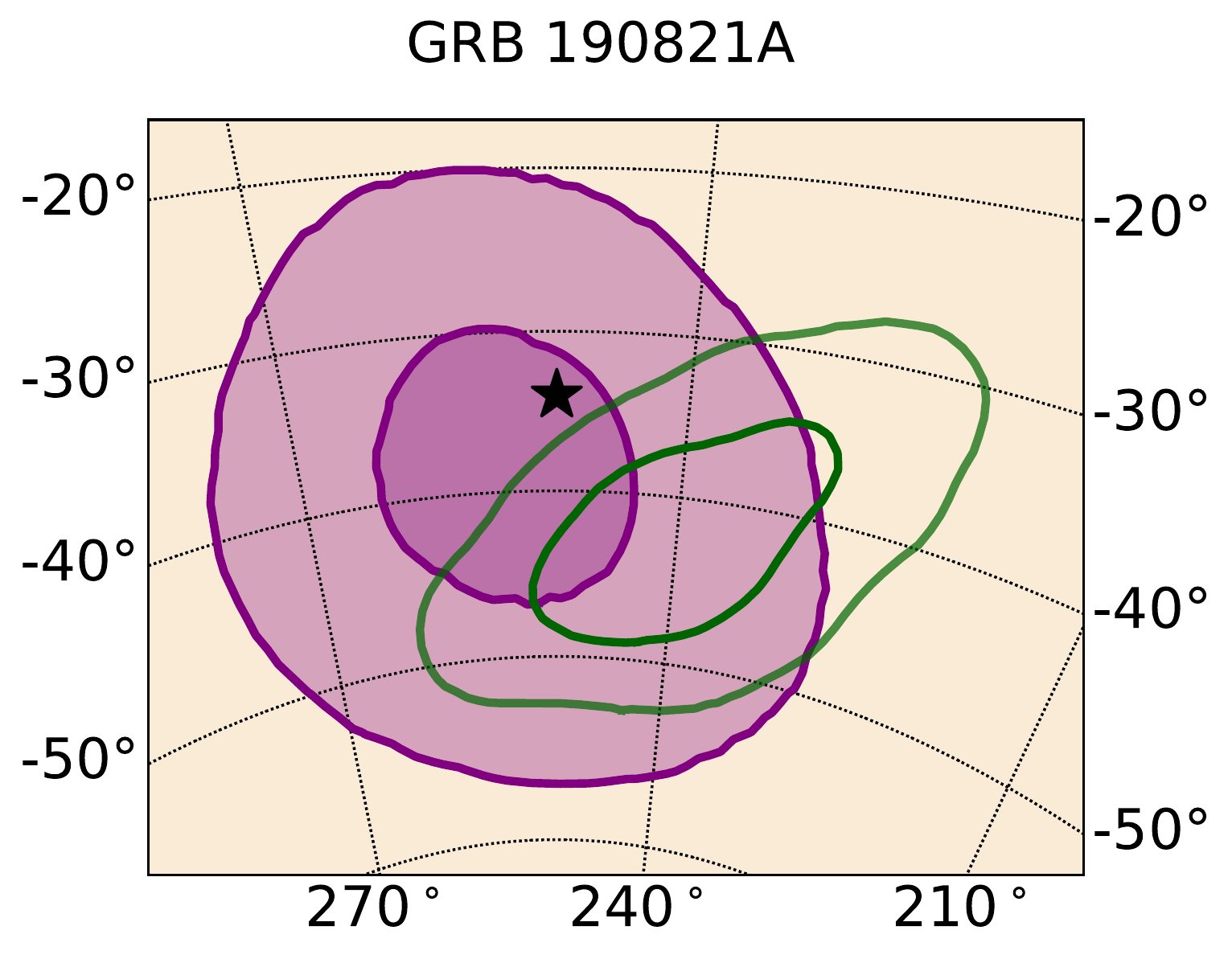}	
		\includegraphics[width=7.5cm]{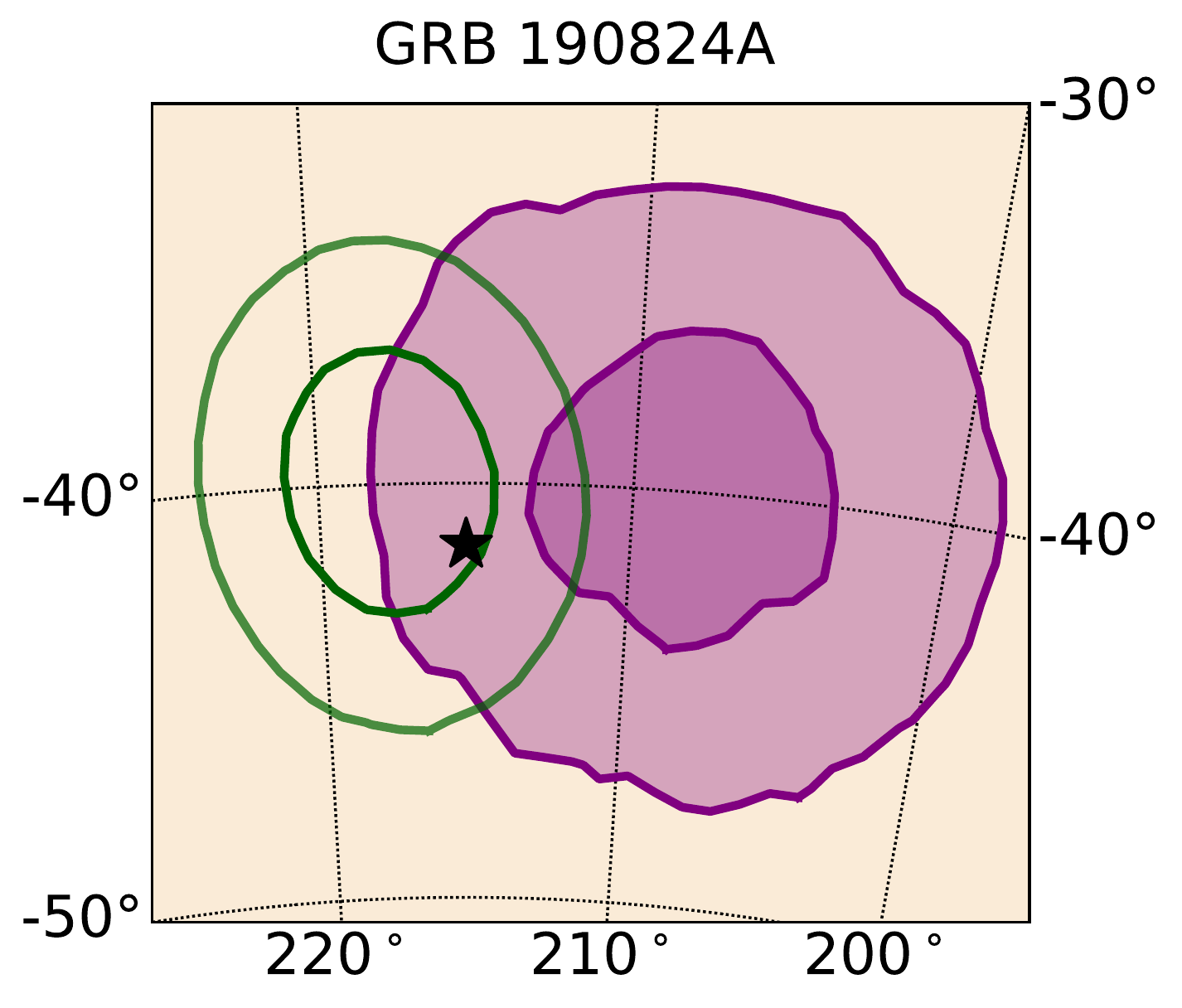}\\		
		\includegraphics[width=15cm]{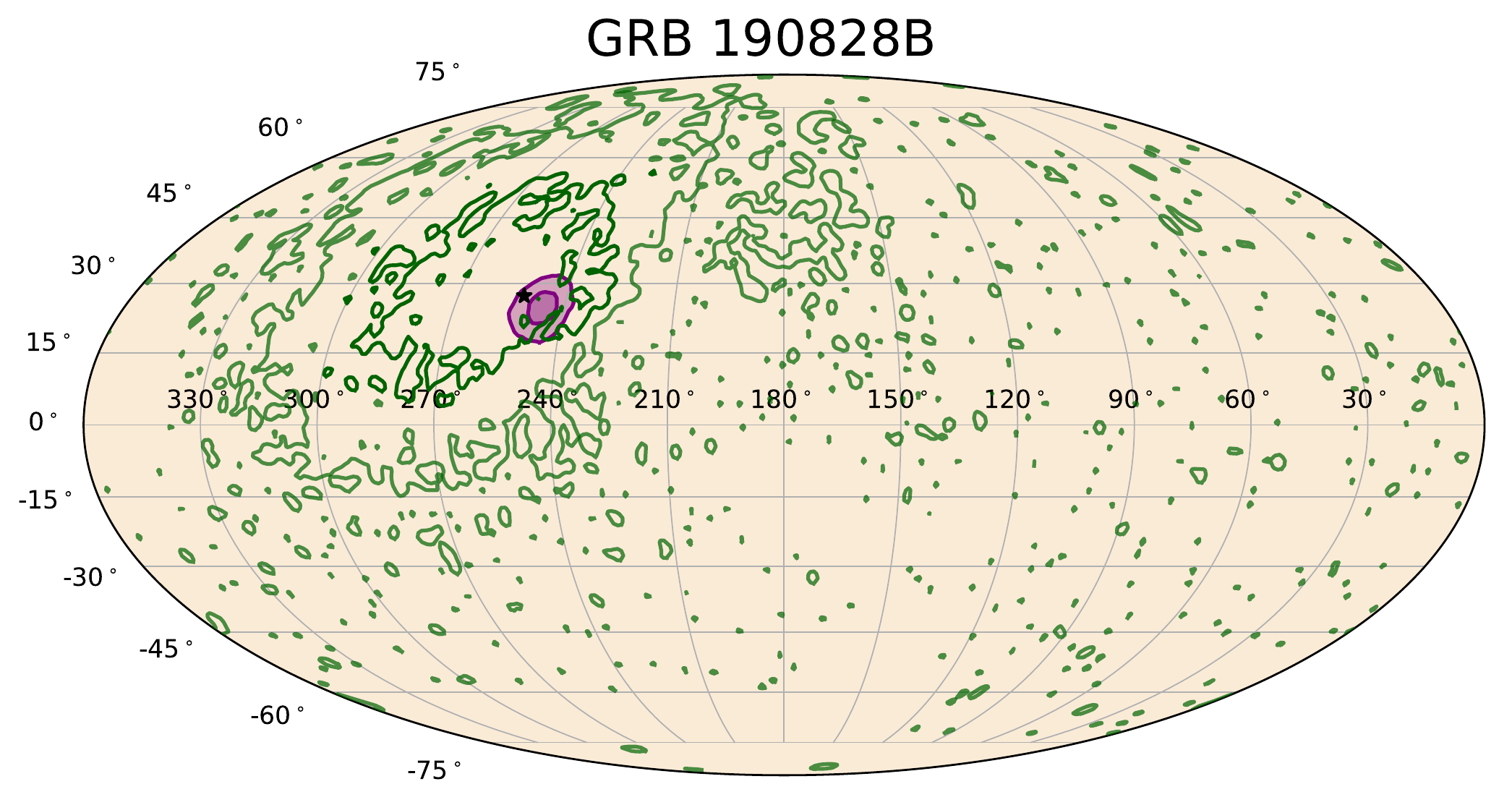}\\	
		\includegraphics[width=7.5cm]{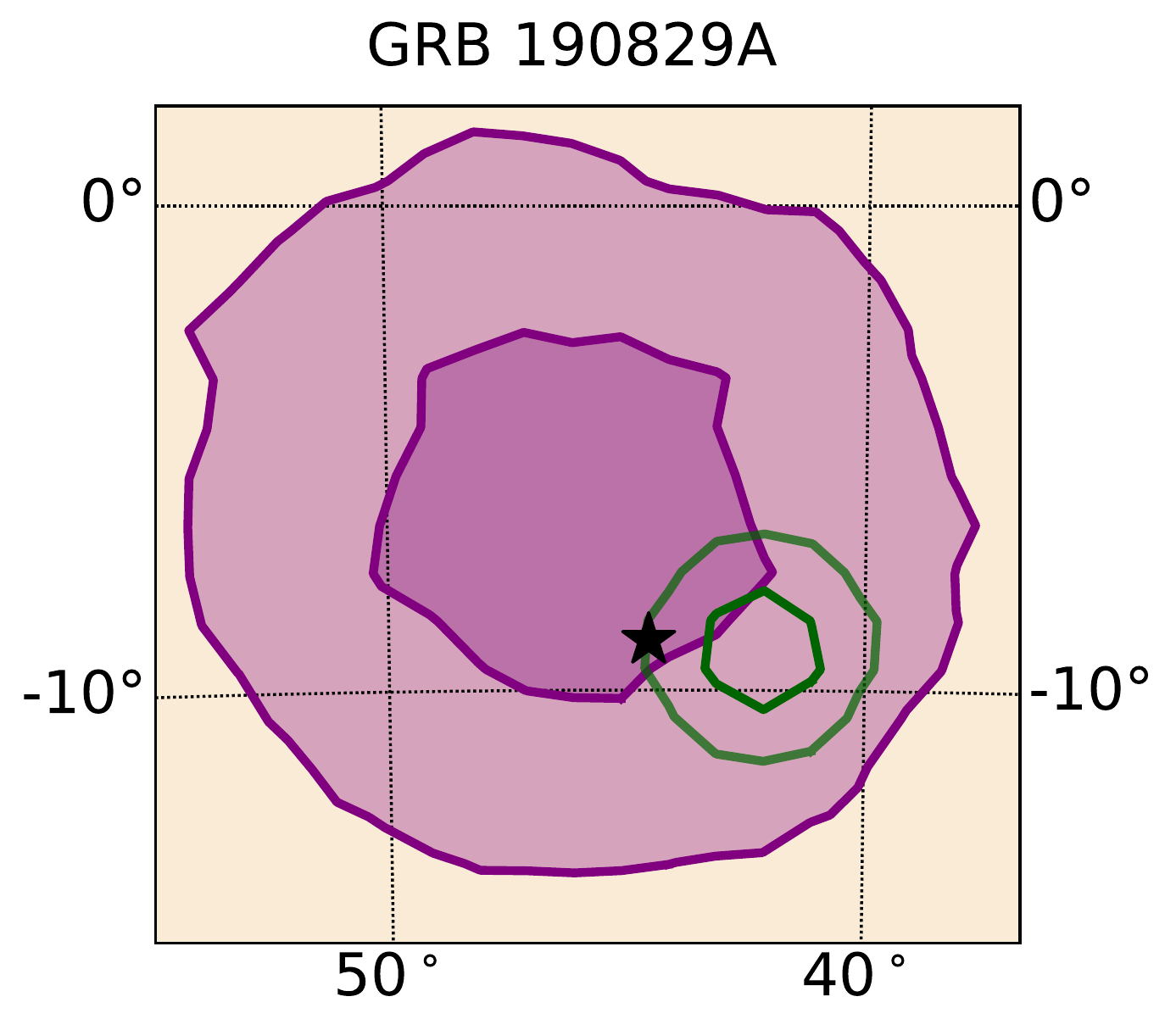}	
	\end{center}
\end{figure}

%%References

\end{document}